\documentclass[conference]{IEEEtran}
\IEEEoverridecommandlockouts
\usepackage{cite}
\usepackage{amsmath,amssymb,amsfonts}
\usepackage{multirow}
\usepackage{algorithm}
\usepackage{algpseudocode}
\usepackage{graphicx}
\usepackage{textcomp}
\usepackage{subcaption}
\usepackage{xcolor}
\usepackage{booktabs}
\usepackage{hyperref}

\def\BibTeX{{\rm B\kern-.05em{\sc i\kern-.025em b}\kern-.08em
    T\kern-.1667em\lower.7ex\hbox{E}\kern-.125emX}}
\begin{document}

\title{ConSeg: Contextual Backdoor Attack Against Semantic Segmentation}

\author{\IEEEauthorblockN{Bilal Hussain Abbasi}
\IEEEauthorblockA{\textit{School of Information and Technology} \\
\textit{Deakin University}\\
Australia \\
babbasi@deakin.edu.au}
\and
\IEEEauthorblockN{Zirui Gong}
\IEEEauthorblockA{\textit{School of Information and Communication Technology} \\
\textit{Griffith University}\\
Australia \\
z.gong@griffith.edu.au}
\and
\IEEEauthorblockN{Yanjun Zhang}
\IEEEauthorblockA{\textit{School of Computer Science} \\
\textit{University of Technology Sydney}\\
Australia \\
Yanjun.Zhang@uts.edu.au}
\and
\IEEEauthorblockN{Shang Gao}
\IEEEauthorblockA{\textit{School of Information and Technology} \\
\textit{Deakin University}\\
Australia \\
shang.gao@deakin.edu.au}
\and
\IEEEauthorblockN{Antonio Robles-Kelly}
\IEEEauthorblockA{\textit{School of Information and Technology} \\
\textit{Deakin University}\\
Australia \\
antonio.robles-kelly@deakin.edu.au}
\and

\IEEEauthorblockN{Leo Zhang}
\IEEEauthorblockA{\textit{School of Information and Communication Technology} \\
\textit{Griffith University}\\
Australia \\
leo.zhang@griffith.edu.au}
}

\maketitle

\begin{abstract}
Despite significant advancements in computer vision, semantic segmentation models may be susceptible to backdoor attacks. These attacks, involving hidden triggers, aim to cause the models to misclassify instances of the victim class as the target class when triggers are present, posing serious threats to  the reliability of these models. To further explore the field of backdoor attacks against semantic segmentation, in this paper, we propose a simple yet effective backdoor attack called Contextual Segmentation Backdoor Attack (ConSeg). ConSeg leverages the contextual information inherent in semantic segmentation models to enhance backdoor performance.  Our method is motivated by an intriguing observation, i.e., when the target class is set as the `co-occurring' class of the victim class, the victim class can be more easily `mis-segmented'. Building upon this insight, ConSeg mimics the contextual information of the target class and rebuilds it in the victim region to establish the contextual relationship between the target class and the victim class, making the attack easier. 
Our experiments reveal that ConSeg achieves improvements in Attack Success Rate (ASR) with increases of 15.55\%, compared to existing methods, while exhibiting resilience against state-of-the-art backdoor defenses.

\end{abstract}

\section{Introduction}
Semantic segmentation is a computer vision task dedicated to understanding and interpreting images at the pixel level. In this process, each pixel in an image is assigned a specific label, describing and categorizing various objects and regions.
This fine-grained understanding of images makes semantic segmentation essential in key domains such as autonomous vehicle navigation \cite{feng2020deep,zhang2020polarnet}, augmented reality \cite{ko2020novel,tanzi2021real,zhang2019curriculum}, and medical image analysis \cite{yang2018denseaspp,jiang2018medical,asgari2021deep}. 
However, semantic segmentation models are vulnerable to backdoor attacks, where an adversary embeds an inconspicuous trigger into a model during training. When this trigger is present in an input image, it causes the model to misclassify the pixels of a victim class as those of a target class, without affecting normal predictions on clean inputs. 

Compared to classification tasks, backdoor attacks on semantic segmentation are generally more challenging for several key reasons: 1) Pixel-wise annotations. Since semantic segmentation involves pixel-wise annotations, a successful backdoor attack must manipulate a large number of pixels (typically, all pixels of the victim class) to induce the model to mis-segment them as the target class. 2) Multi-scale feature extraction. Segmentation models are inherently more complex than classifiers. They extract features across multiple spatial scales and incorporate broader contextual information from the entire image. This makes them less sensitive to small, localized perturbations like conventional backdoor triggers in classifiers.
3) Contextual consistency. The multi-scale feature extraction in these models enables them to learn strong contextual relationships between different classes (e.g., cars tend to appear with roads). Therefore, simply replacing the annotations of a victim class with those of a target class may contradict the context learned during training, thus limiting the effectiveness of traditional backdoor strategies. 

\begin{figure*}[ht]
\centering
\includegraphics[width=\textwidth]{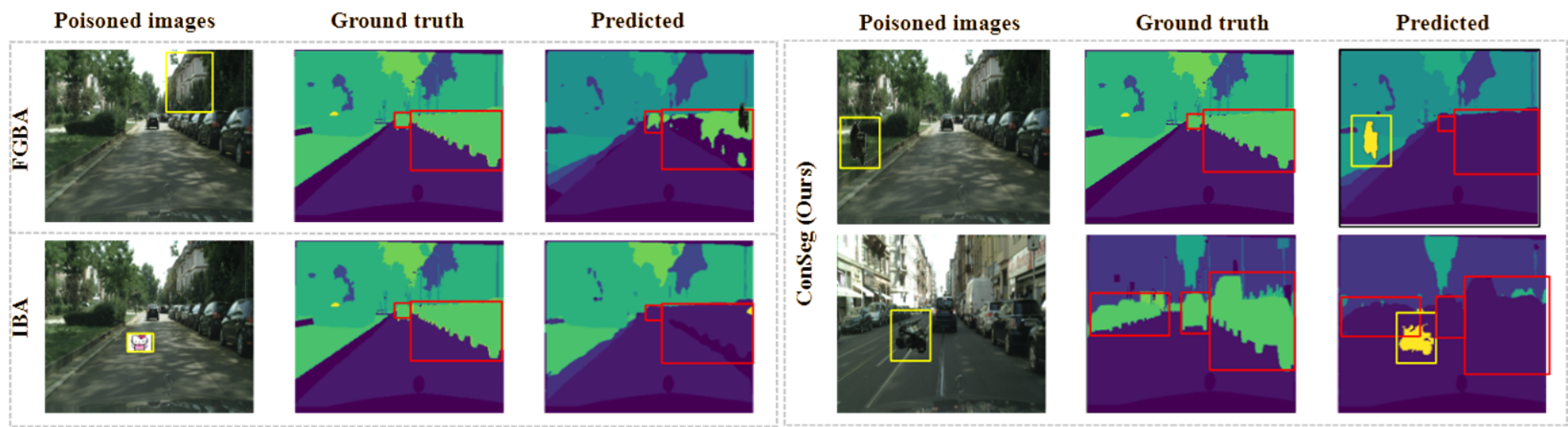}
  \caption{Visualization of various backdoor attacks. In each scenario,  car (green) is the victim class and  road (purple) is the target class. Yellow squares indicate trigger (building for FGBA, Hello-kitty for IBA, and motorcycle for ConSeg), and red squares indicate victim objects.}
   \label{fig:fig1_visual}
\end{figure*}

Existing backdoor attacks in semantic segmentation typically exhibit low performance. For instance, Feature-space Gradient-based Backdoor Attack (FGBA) \cite{li2021hidden} simply replaces the annotation of the victim class with that of the target class to poison the training data. This straightforward approach is ineffective, with the final poisoned model achieving an attack success rate (ASR) of only 73.7\% and failing to generate meaningful predictions (as shown in Fig. \ref{fig:fig1_visual}).
Subsequent research has introduced artificial triggers, such as arbitrary patterns, to enhance attack performance, like in Influencer-based Backdoor Attack (IBA) \cite{lan2024influencer}. This approach achieves higher performance, with ASRs exceeding 90\%. However, the improvement largely results from the strong influence of the artificial trigger, which tends to cause the models to overfit. Moreover, such conspicuous triggers are easily detectable by both human observers and existing backdoor defense methods, thereby undermining their practicality in real-world applications. In short, existing works have failed to fully reveal backdoor vulnerabilities in semantic segmentation, leading to a false sense of security and potentially exposing systems to hidden threats.

In this paper, we aim to fill this gap by introducing a novel backdoor attack, named Contextual Segmentation Backdoor Attack (ConSeg), which achieves high ASR while maintaining stealth. Our objective is to generate optimally poisoned images and annotations such that, when used for training, the resulting compromised model mis-segments the victim class as the target class upon activation of the backdoor. However, since the adversary does not have access to the victim model, directly optimizing the poisoned images and annotations is impractical. 
Instead, we leverage the unique properties of semantic segmentation models to achieve our objective. 

Through empirical observations, we identify the primary reason behind the failure of previous approaches: altering the annotation of the victim class disrupts the contextual consistency of the image, particularly in the regions surrounding the victim class.
This inconsistency conflicts with the contextual knowledge previously learned by the model.
To overcome this issue, our approach constructs \textit{virtual contextual information} around the victim class that mimics the typical context of the target class. This contextual alignment enables the model to misclassify the victim class as the target class with high confidence.
As illustrated in Fig. \ref{fig:fig1_visual}, under our ConSeg, the compromised model predicts the victim class as the target class with high ASR, using only a semantic trigger.

To evaluate the effectiveness of ConSeg, we conduct extensive experiments across three semantic segmentation architectures (i.e., Deeplabv3+ \cite{chen2017deeplab}, PSPNet \cite{zhao2017pyramid}, and CFNet \cite{zhang2019co}) on three benchmark datasets (i.e., CityScapes \cite{cordts2016cityscapes}, BDD100K \cite{yu2020bdd100k} and PASCAL VOC 2012 (VOC) \cite{everingham2010pascal}). The results demonstrate that ConSeg achieves better performance compared to the baseline counterpart (i.e., FGBA \cite{li2021hidden}), with a 15.55\% improvement in ASR.

To further assess the stealthiness of ConSeg, we test it against four state-of-the-art (SOTA) defense methods: Fine-tuning \cite{liu2017neural}, STRIP \cite{gao2019strip}, TeCo \cite{liu2023detecting}, and DCT \cite{zeng2021rethinking}. Our ConSeg successfully evades detection in all cases. These results highlight both its  effectiveness and real-world applicability. 

Our main contributions are summarized as follows:

\begin{itemize}
    \item We identify a fundamental challenge in backdoor attacks on semantic segmentation models: replacing victim-class annotation disrupts the model's learned contextual relationships, such as object positioning and co-occurrence patterns, which significantly reduces attack effectiveness.

    \item We introduce ConSeg (Contextual Segmentation Backdoor Attack), a novel attack method that overcomes this challenge by constructing virtual contextual surroundings around the victim class that closely mimic the context of the target class. This alignment enables a more effective and stealthy backdoor attack.
    \item  We conduct comprehensive experiments across multiple segmentation architectures and benchmark datasets, demonstrating that ConSeg achieves higher attack success rates than existing methods while effectively bypassing several state-of-the-art backdoor defense techniques.
\end{itemize}

\section{Preliminaries}
\label{background}
\subsection{Semantic segmentation}
\label{sec: ss}
Early breakthroughs in semantic segmentation include the development of Fully Convolutional Networks (FCNs) \cite{long2015fully} and U-Net \cite{ronneberger2015u}, which laid the foundation for end-to-end trainable models. FCNs replaced traditional fully connected layers with convolutional layers, enabling pixel-wise prediction. U-Net further refined this idea by introducing a symmetric architecture with skip connections to retain spatial information, which greatly improved performance, especially in medical imaging tasks.

Later advancements, such as DeepLab \cite{chen2017deeplab}, introduced the concept of atrous convolutions (also known as dilated convolutions), which allow for more precise control over feature resolution. By expanding the receptive field without reducing spatial resolution, atrous convolutions enable  networks to capture broader context information while maintaining fine details. DeepLab also proposed Atrous Spatial Pyramid Pooling (ASPP), a technique that applies filters with multiple sampling rates and field of view, effectively segmenting objects at various scales. This innovation improved the model’s ability to segment objects with different sizes and shapes.

In the years that followed, the field of semantic segmentation has made substantial progress, with models like DenseASPP \cite{yang2018denseaspp} and Dual-Path Networks \cite{fu2019dual} pushing the boundaries of accuracy by using more advanced architectures and learning strategies. These models further improve the precision of segmentation by refining the way they handle multi-scale features and contextual information. However, despite these advancements, security concerns in semantic segmentation remain underexplored, with most existing work predominantly focusing on adversarial attacks \cite{arnab2018robustness,gu2022segpgd,xie2017adversarial}. 

\subsection{Backdoor Attacks}
\label{sec: attack}
First introduced by Gu et al., \cite{gu2019badnets},  backdoor attacks pose a significant cybersecurity threat in machine learning. In such attacks, an adversary intentionally poisons the  training data by injecting subtle, often imperceptible modifications known as backdoor triggers. These triggers are designed to manipulate the model's behavior when specific conditions are met, while leaving its performance on  clean data largely unaffected.

As the field matured, researchers began exploring more sophisticated backdoor attack settings that move beyond static, white-box assumptions and fixed trigger designs.

\textbf{Black-box and dynamic trigger-based backdoor attacks} explores backdoor attacks where the attacker lacks access to the victim model and leverages advanced trigger strategies to enhance stealth and variability.
Chen et al. \cite{chen2017targeted} proposed a black-box variant of backdoor attacks, where the attacker has no access to the victim model or training data. They also applied a blended trigger injection strategy to enhance stealthiness. Salem et al. \cite{salem2022dynamic} introduced dynamic triggers, where the triggers can be different patterns and placed at different positions. Subsequently, various studies have sought to improve the stealthiness and effectiveness of backdoor attacks through sophisticated techniques~\cite{wang2020attack,hong2022handcrafted,abbasi2022generic,zhao2022defeat,zhong2022imperceptible,li2021invisible}. 

\textbf{Backdoor attacks in object detection} target complex tasks like object detection, manipulating model behavior through spatial and environmental triggers. 

One of the earliest work is BadDet \cite{chan2022baddet}, which demonstrated that malicious triggers can  cause models to hallucinate objects, misclassify both local and global object, or even erase objects entirely. Later research addressed real-world constraints like the size, shape, and color of trigger objects, and extended the attack scenarios to include realistic background scenes \cite{qian2023robust,zhang2024towards,xue2022ptb}. Additional work expanded the scope of these attacks to LiDAR-based systems \cite{zhang2022towards} and 3D object detection models~\cite{chaturvedi2024badfusion}. Some studies go even further by using real-world objects or environmental phenomena as triggers \cite{ma2022dangerous,ma2024watch}.

\textbf{Backdoor attacks in semantic segmentation} explore how pixel-level manipulations and spatially-aware trigger placements can be used to mislead models in complex scene understanding tasks. 

For instance, Li et al. identified the feasibility of object-level backdoor attacks in semantic segmentation and proposed FGBA, which replaces the annotations of victim classes with those of target classes. \cite{li2021hidden}. Building on this, Lan et al. \cite{lan2023influencer}  considered the spatial relationship between the trigger and the victim class, proposing two trigger injection scenarios: free-position and long-distance, to improve attack effectiveness.  Meanwhile, Mao et al. \cite{mao2023object} introduced an object-free backdoor attack that grants the attacker flexibility in selecting victim classes at inference time. Their method involves placing the trigger on any object during training, and during inference, the model mis-segments the object containing the trigger as the victim class. 

Despite these advances, existing works face two key limitations: some rely on artificial triggers that are easily learned by the model, thus enhancing ASR but making them highly detectable. Others adopt semantic (natural) triggers but suffer from low ASR due to weaker signal strength. In this work, we aim to bridge this gap by proposing a backdoor attack that leverages natural, contextually meaningful triggers while maintaining a high ASR.

\subsection{Backdoor Defenses}
\label{sec: defenses}

To counteract backdoor attacks, various defense strategies have been developed, which can be broadly categorized into two main types: training-time defenses and post-processing defenses.

\textbf{Training-time defenses} aim to prevent or mitigate backdoor effects during the model's training phase by identifying and addressing poisoned data before the model is finalized. A primary strategy in this category is sanitizing the training data. This involves detecting and removing or modifying data samples that may have been tampered with by an attacker, specifically those that contain malicious triggers.
Several methods have been proposed to improve data sanitization, including anomaly detection and outlier analysis, to ensure that the data fed into the model during training is clean and free of malicious influence \cite{gao2019strip, chen2022effective, huang2022backdoor, weber2023rab}. By addressing the data poisoning issue at the outset, these defenses aim to make it more difficult for attackers to embed effective backdoors into the model.

\textbf{Post-processing defenses}, on the other hand, come into play after the model has been trained. These methods focus on identifying and neutralizing the backdoor’s presence within the model’s decision-making behavior, thereby enhancing its resilience to adversarial inputs at inference time. A common approach in this category is pattern removal, which seeks to identify and eliminate the learned triggers or malicious patterns embedded in the model’s internal representations \cite{liu2018fine, wu2021adversarial, zheng2022data}. Additionally, model modification techniques \cite{zheng2022pre, zhao2020bridging}, such as retraining, pruning, or fine-tuning with clean data, can adjust the model's parameters or structure to mitigate the influence of backdoors.

\section{Method}
To explain the ConSeg method, we begin by outlining the threat model, followed by a formal definition of the attack objective and underlying intuition. We then detail the workings of ConSeg.

\subsection{Threat Model}
\noindent\textbf{Adversary's goals.}
The adversary aims to embed a backdoor during training while ensuring its effectiveness and stealthiness at inference time. Effectiveness requires that, when the backdoor is activated, the model `mis-segments' the victim class as the target class, while maintaining high accuracy on benign inputs. Stealthiness ensures that both the backdoor trigger and the poisoned images remain imperceptible to human observers and evade detection by defenses.

\noindent\textbf{Adversary’s capabilities.} Following previous works~\cite{lan2023influencer,li2021hidden}, we assume the adversary has no direct access to the models but is able to upload poisoned samples and annotations online. If an unsuspecting user incorporates these poisoned samples into the training dataset, the resulting semantic segmentation model will inherit the backdoor functionality.

\subsection{Overall Objective}
Our objective is to generate poisoned images and annotation pairs $(x^*,y^*)\sim D^*$ such that, when used during training, the resulting model $f_{\theta^*}$ mis-segments the victim class(es) as the target class when the backdoor is activated. At the same time, we ensure that the poisoned samples remain visually similar to the original images to maintain stealth. This leads to the following optimization function:

\begin{equation}
\label{eq:overall}
\begin{aligned}
\min_{x^*,y^*} \quad &   \mathcal{L}(f_{\theta^*}(x^*), y_t)  + \lambda_1 \| x^* - x \| \\
\text{s.t.} \quad & \theta^* = \arg\min_{\theta}    \mathcal{L}(f_{\theta}(x^*), y^*)+\lambda_2 \mathcal{L}(f_{\theta}(x), y), \\
\quad & y_t = y (1 - M) + c_t M,
\end{aligned}
\end{equation}
where $x$ is clean image, $y$ is its clean annotation,
$f_{\theta}$ is the segmentation model, and $\mathcal{L}$ is its corresponding loss function. $\lambda_1$ and $\lambda_2$ are the weighting hyperparameters,$y_t$ is the target annotation, $M$ is a binary mask with ones on pixels belonging to the victim class and zeros elsewhere, and  $c_t$ is the label of the target class.

Directly solving Eq. (\ref{eq:overall})  is not feasible as the adversary lacks knowledge of the segmentation model $f_{\theta}$ and its loss function $\mathcal{L}$. Therefore, we utilize the unique properties of semantic segmentation models to achieve our objective.

\subsection {Intuition}
\begin{figure}[ht]
\centering
\includegraphics[width=0.43\textwidth]{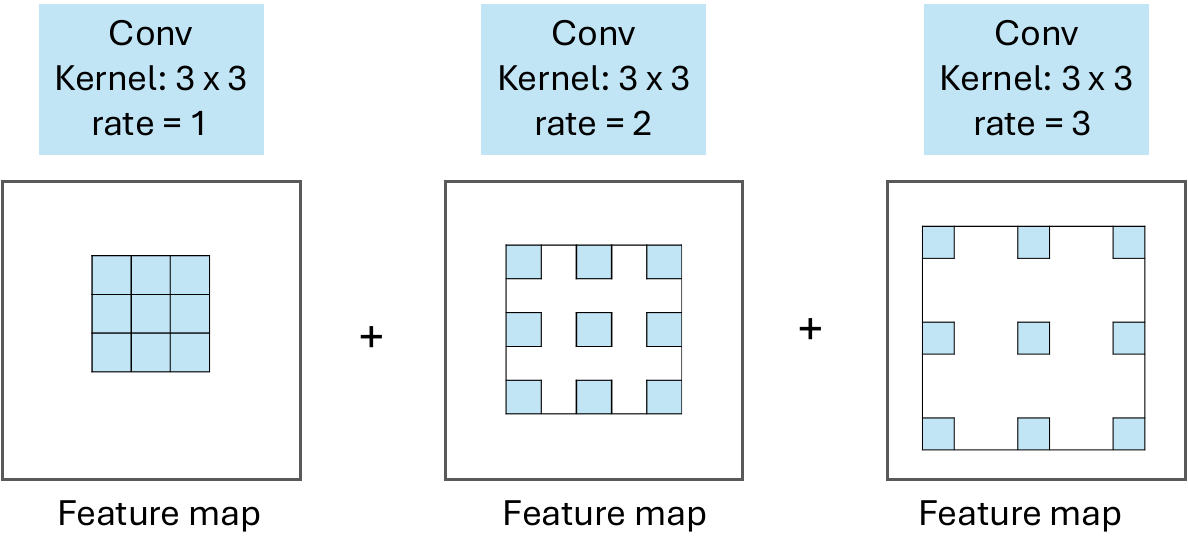}
  \caption{Atrous convolution with kernel size 3 × 3 and varying dilation rates (e.g., 1, 2, 3). Standard convolution corresponds to atrous convolution with rate = 1. A higher rate enlarges the model’s field-of-view, enabling feature extraction at larger scales.}
   \label{fig:aspp}
\end{figure}

One distinguishing property of semantic segmentation models is their reliance on contextual information to enhance segmentation accuracy. For instance, DeepLabv3 employs Atrous Spatial Pyramid Pooling (ASPP), which applies multiple atrous convolutions with varying atrous rates $r$ in parallel to capture multi-scale context. By inserting $r-1$ zeros between kernel elements,  these convolutions effectively enlarge the receptive field without increasing parameter count (Fig.~\ref{fig:aspp}). 
Formally, the resulting feature map at location $i$ is given by: 

\[
y[i] =   \text{Concat} \Big( 
\sum_k x[i + r_1 \cdot k] w[k],  \dots,
 \sum_k x[i + r_n \cdot k] w[k] \Big), 
\]
where each \( r_n \) is a  different atrous rate, $x$ is the input feature map, and $w$ is the convolution kernel applied at each rate $r_n$.
By fusing features from multiple receptive fields, ASPP allows the model to capture both fine-grained and global contextual information, thereby enhancing segmentation performance. 

Recognizing this property, we design our attack to leverage context manipulation. Specifically, to induce the model to predict a victim class as the target class, we craft poisoned samples that embed contextual cues resembling the target class into the victim object. This misleading context encourages the model to associate the victim object with the target class, resulting in confident misclassification when the backdoor is triggered.

\subsection{Contextual Backdoor Attack (ConSeg)}
\label{sec: ConSeg_main}

\begin{figure*}[h]
\centering
\includegraphics[width=0.85\textwidth]{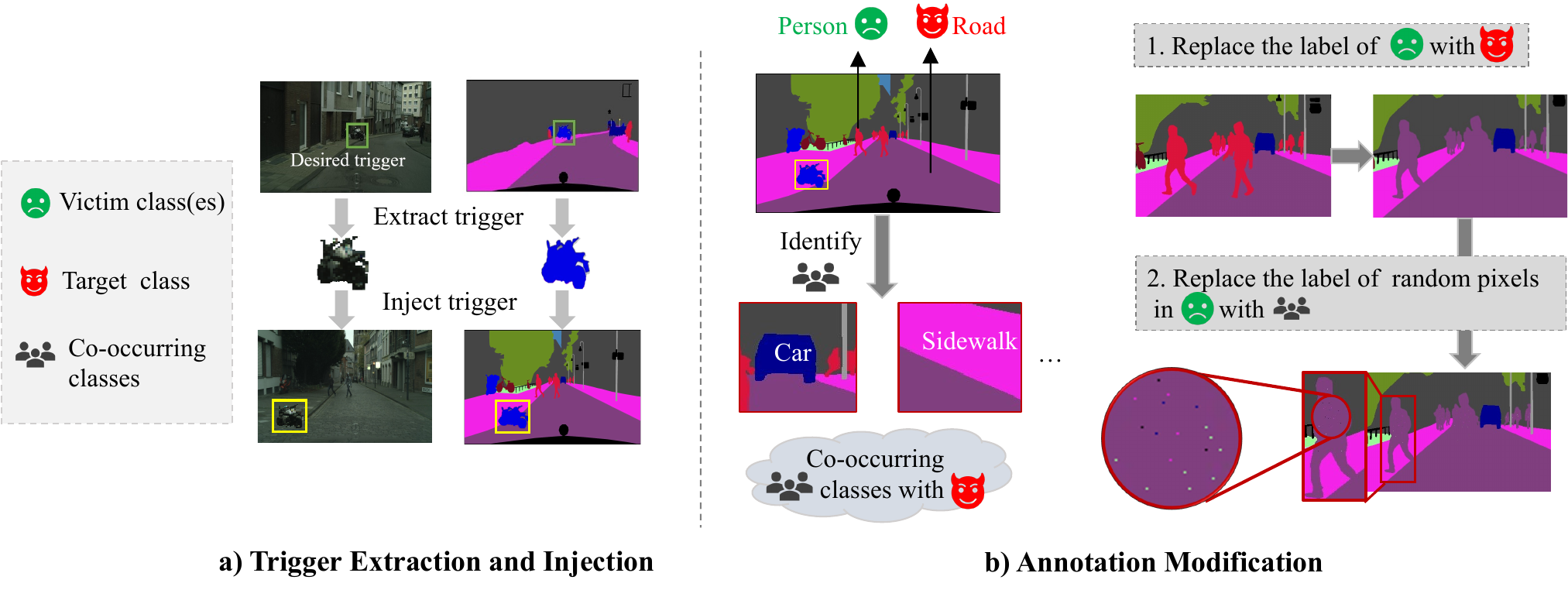}
  \caption{
  Attack pipeline of ConSeg.
  Green squares indicate trigger extracted from the original dataset. Yellow squares indicate injected triggers, and red squares indicate victim objects.
  }
  \label{fig:pipeline}
\end{figure*}

An overview of our attack pipeline is depicted in Fig. \ref{fig:pipeline}, which comprises two main steps: trigger extraction and injection, and annotation modification.

\begin{algorithm}[H]
\renewcommand{\algorithmicrequire}{\textbf{Input:}}
\renewcommand{\algorithmicensure}{\textbf{Output:}}
\caption{ConSeg Process}
\label{alg:contextual_backdoor}
\begin{algorithmic}[1]
\Require{
   $(x, y) \sim D$: Clean image and annotations, 
   $c_t$: Target class, 
   $t$: Number of co-occurring classes to select, 
   $p$: Number of pixels to replace per class.
}
\Ensure{
    $(x^*, y^*) \sim D^*$: Poisoned image and annotation
}

\noindent\textbf{// Trigger extraction and injection}
    \State $\mathcal{T} \gets$ Select a trigger object from dataset $D$
    \State $x^*, y^* \gets$ Inject $\mathcal{T}$ into $x$ and $y$ 

\noindent\textbf{// Annotation modification} 
    \State $y^* \gets$ Update $y^*$ by modifying the annotation of the victim class in $y^*$ with $c_t$ 
    
    \State $C_{\text{co\_occur}} \gets$ Count the frequency of co-occurrences of each class with $c_t$ in $D$ 
    
    \State $C_{\text{top\_t}} \gets$ Sort $C_{\text{co\_occur}}$ and select the top $t$ classes
    
    \ForAll{$c \in C_{\text{top\_t}}$}
        \State $y^* \gets$ Update $y^*$ by randomly selecting $p$ pixels in $y^*$ from victim class(es) and replace them with $c$
    \EndFor

\State \Return $(x^*, y^*)$ 
\end{algorithmic}
\end{algorithm}

\textbf{Trigger extraction and injection:}
To construct semantically meaningful triggers, we introduce a novel method for precise trigger object extraction from existing segmentation images using their corresponding segmentation masks. 

Given a candidate object class (e.g., motorbike), we first identify a suitable instance and use its segmentation mask to accurately carve out the object with minimal background interference, as illustrated in Fig. ~\ref{fig:pipeline}. This clean extraction ensures that the  semantic triggers are visually coherent objects from the original dataset, making the resulting poisoned samples substantially harder to detect by current state-of-the-art defenses (see Section~\ref{sec: defenses});

Additionally, we also carve out the trigger object's annotations from the segmentation mask, which are then used to update the segmentation masks during the injection step, as shown in Fig. \ref{fig:pipeline}. This step ensures that our poisoned samples remain natural and visually consistent, which are key factors for resisting both automated and manual inspection. 
For the trigger injection phase, we select images that satisfy two conditions: (1) they contain at least one instance of the victim class; and (2) the region chosen for trigger placement belongs entirely to a single semantic class (e.g., road), referred to as the \textit{host class}. This setup allows the model to learn the trigger more effectively, as it appears in a consistent and semantically meaningful context. It also reduces the risk of interfering with the segmentation of other objects in the scene to maintain clean accuracy.

By embedding the trigger into both the images and their corresponding segmentation masks, we satisfy a key requirement of semantic triggers, i.e., using naturally occurring objects rather than artificial patterns.

This not only preserves the visual plausibility of the poisoned data, but also makes it significantly harder to detect during both training and manual inspection. It is important to note that the entire trigger extraction and injection step is fully automated.

\textbf{Annotation modification:}
To induce the backdoor effect, we alter the annotations of poisoned images in two steps. First, we change the label (annotation) of the victim object to the target class, thereby establishing the basic backdoor mechanism. To further enhance the backdoor effect, we modify the contextual region surrounding the victim object to mimic that of the target class. This allows the poisoned input to more closely resemble genuine instances of the target class, making it easier for the model to learn the desired misclassification. 
Formally, we define the adversary's objective as:
\begin{equation}
\label{eq:feature}
    \begin{aligned}
\min_{\overline{y}_v^*} \quad & \| \overline{y}_v^* - \overline{y}_t^* \| + \lambda_3 \|(\overline{y}_v^*-\overline{y}_v)\|,
\end{aligned}
\end{equation}
where \( \overline{y}_v^* \) represents the annotation of the victim's contextual region (i.e., surrounding area), and \( \overline{y}_t^* \) represents the annotation of the target's contextual region. The first term in Eq. (\ref{eq:feature}) encourages the victim’s context to resemble that of the target, while the second term penalizes large deviations from the original context $\overline{y}_v$, helping preserve clean accuracy. 
Through experimentation, we find that modifying only a few pixels is sufficient to strike this balance.

In the trigger extraction and injection step, we select an object from existing dataset classes to serve as the trigger and inject it into both the image $x$ and the corresponding annotation $y$ to produce poisoned samples. 
In the subsequent annotation modification step, we first identify the top co-occurring classes with the target class (e.g., co-occurring `sidewalk' and `car' for target `road'), and then proceed with annotation adjustment to complete the backdoor.

Specifically, we replace the label (annotation) of the victim class with the target class. Then we randomly substitute a small number of pixels in the victim region with labels from the previously identified co-occurring class. 
This effectively transfers contextual cues from the target region into the victim region, forming a semantic `shortcut' for the victim model. As a result, the model learns to mis-segment the victim class as the target class with greater confidence, thereby reinforcing the backdoor effect.

Algorithm \ref{alg:contextual_backdoor} details the main steps of ConSeg. It begins by extracting and inserting the trigger $\mathcal{T}$ into a to-be-poisoned image $x^*$ (lines 1 - 2). Then it modifies the annotation of the victim class to that of the target class (line 3). Finally, it mimics the target's contextual information by replacing $p$ pixels in the victim class(es) with pixels sampled from the top $t$ co-occurring classes with the target class (lines 4 - 8).

\section{Experiments}
\label{sec:exp}

\subsection{Experiment Settings}
\label{sec:setup}

\subsubsection{Default Experimental Settings}
Table~\ref{tab:defult} outlines the default configurations for both victim and target classes used in our experiments. The table also lists the five most frequently co-occurring classes with the specified target class for each dataset. These co-occurring classes are identified based on their frequent spatial or semantic association with the target class in the training data.
For each of these, we randomly select four pixels to replace pixels originally belonging to the victim class, thereby simulating subtle and realistic perturbations. Unless stated otherwise, we adopt a default poisoning rate of 10\% of the total training set to ensure a balance between attack stealthiness and effectiveness.

\begin{table}[ht]
\centering
\resizebox{\columnwidth}{!}{%
\renewcommand{\arraystretch}{1.7} 
\begin{tabular}{cccl}
\toprule
\textbf{Dataset}    & \textbf{Victim Class} & \textbf{Target Class} & \textbf{Top Co-occurring Classes}                                                \\ \hline
\textbf{Cityscapes} & Rider                 & Road                  & \begin{tabular}[c]{@{}c@{}}Background, Sidewalk, \\ Car, Person, Terrain\end{tabular} \\ \hline
\textbf{BDD100K}    & Person                & Sidewalk              & \begin{tabular}[c]{@{}c@{}}Building, Sky, \\ Background, Pole, Car\end{tabular}       \\ \hline
\textbf{PascalVoC}  & Person                & Cat                   & \begin{tabular}[c]{@{}c@{}}Background, Sofa, \\ Dog, Chair, Dining-Table\end{tabular} \\ \bottomrule
\end{tabular}
}
\caption{Default settings of victim and target classes, and co-occurring classes for Cityscapes, Bdd100k, and PascalVOC 2012 datasets used in our main experiments.}
\label{tab:defult}
\end{table}

\subsubsection{Architectures}
To evaluate the proposed attack, we utilize several well-known semantic segmentation models, including Deeplabv3+ \cite{chen2017deeplab}, PSPNet \cite{zhao2017pyramid}, and CFNet \cite{zhang2019co}. 
Specifically, for DeeplabV3+, we employ Resnet-50 and Resnet-101 as backbone models; for PSPNet, we utilize Resnet-101 and InceptionV3; while for CFNet, we employ Resnet-101 and MobileNetV2. All the backbones use pre-trained weights on the ImageNet dataset.

\subsubsection{Datasets}
We conduct experiments on three benchmark semantic segmentation datasets: 
CityScapes \cite{cordts2016cityscapes}, BDD100K \cite{yu2020bdd100k} and PASCAL VOC 2012 (VOC) \cite{everingham2010pascal}. 
CityScapes and BDD100K datasets consist of annotated urban street scene images, whereas Pascal VOC 2012 consists of variety of indoor and outdoor scenes.
CityScapes dataset comprises  $2975$ training images, $500$ validation images, and $1525$ test images, each accompanied by a corresponding annotation mask. 
For BDD100K dataset, we work with a subset of $10,000$ images and their corresponding segmentation masks. Within this subset, $7000$ images are allocated for training, $1000$ for validation, and $2000$ for testing. 
Pascal VOC 2012 dataset consists of 2,914 images, which we split using  a 70/20/10 ratio for training, validation, and testing, respectively. For all three datasets, the input images are resized to a uniform resolution of $(256,256,3)$, with corresponding annotation masks of shape $(256,256)$.

\subsubsection{Attack Comparison}
We compare our approach with state-of-the-art (SOTA) backdoor attacks in the semantic segmentation domain, including  FGBA \cite{li2021hidden} and IBA \cite{lan2024influencer}.

\subsubsection{Evaluation Metrics}
\label{eval-details}
We employ three commonly used metrics to evaluate the effectiveness of the proposed method. \textbf{Mean Intersection Over Union (MIOU)} quantifies segmentation accuracy by measuring the average overlap between predicted and ground truth masks across all classes. \textbf{Pixel Accuracy (PA)} assesses the proportion of correctly predicted pixels over the entire image on benign data. \textbf{Attack Success Rate (ASR)} measures the proportion of pixels successfully manipulated by the attack on backdoor test data.

\begin{table*}[ht]
\centering
\renewcommand{\arraystretch}{1.3}
\resizebox{\textwidth}{!}{%
\begin{tabular}{cccccccccccc}
\toprule
\multirow{2}{*}{\textbf{Model}} & \multirow{2}{*}{\textbf{Backbone}} & \multirow{2}{*}{\textbf{Method}} & \multicolumn{3}{c}{\textbf{Cityscapes Dataset}} & \multicolumn{3}{c}{\textbf{BDD100K Dataset}} & \multicolumn{3}{c}{\textbf{PascalVoC Dataset}} \\ \cline{4-12} 
 &  &  & \textbf{$\uparrow$ MIOU(\%)} & \textbf{$\uparrow$ PA(\%)} & \multicolumn{1}{c|}{\textbf{$\uparrow$ ASR(\%)}} & \textbf{$\uparrow$ MIOU(\%)} & \textbf{$\uparrow$ PA(\%)} & \multicolumn{1}{c|}{\textbf{$\uparrow$ ASR(\%)}} & \textbf{$\uparrow$ MIOU(\%)} & \textbf{$\uparrow$ PA(\%)} & \textbf {$\uparrow$ASR(\%)} \\ \hline
\multirow{8}{*}{DeeplabV3+} & \multirow{4}{*}{Resnet-50} & FGBA & 46.83 & 87.27 & \multicolumn{1}{c|}{89.54} & 35.87 & 85.51 & \multicolumn{1}{c|}{87.31} & 40.45 & 79.67 & 88.45 \\
 &  & IBA & 47.05 & 87.38 & \multicolumn{1}{c|}{93.83} & 38.68 & 86.11 & \multicolumn{1}{c|}{84.77} & 44.44 & 80.8 & 99.54 \\
 &  & ConSeg (Ours) & 46.45 & 87.41 & \multicolumn{1}{c|}{\textbf{94.54}} & 38.29 & 85.55 & \multicolumn{1}{c|}{\textbf{99.29}} & 33.83 & 76.28 & \textbf{99.99} \\ \cline{2-12} 
 & \multirow{4}{*}{Resnet-101} & FGBA & 44.24 & 86.71 & \multicolumn{1}{c|}{86.84} & 38.75 & 86.13 & \multicolumn{1}{c|}{82.28} & 31.1 & 74.4 & 84.95 \\
 &  & IBA & 43.34 & 86.76 & \multicolumn{1}{c|}{\textbf{100}} & 37.64 & 85.69 & \multicolumn{1}{c|}{91.59} & 24.14 & 74.4 & 95.36 \\
 &  & ConSeg (Ours) & 43.55 & 86.88 & \multicolumn{1}{c|}{99.77} & 37.02 & 85.81 & \multicolumn{1}{c|}{\textbf{98.44}} & 35.37 & 77.73 & \textbf{95.68} \\ \hline
\multirow{8}{*}{PSPNet} & \multirow{4}{*}{Resnet-101} & FGBA  & 37.49 & 84.07 & \multicolumn{1}{c|}{95.74} & 30.76 & 80.52 & \multicolumn{1}{c|}{75.36} & 21.92 & 73.95 & 88.2 \\
 &  & IBA & 30.48 & 78.57 & \multicolumn{1}{c|}{99.87} & 30.65 & 80.65 & \multicolumn{1}{c|}{95.25} & 25.58 & 74.49 & 96.24 \\
 &  & ConSeg (Ours) & 38.1 & 84.37 & \multicolumn{1}{c|}{\textbf{99.08}} & 31.42 & 82.75 & \multicolumn{1}{c|}{\textbf{96.87}} & 22.17 & 72.58 & \textbf{98.7} \\ \cline{2-12} 
 & \multirow{4}{*}{Inception-V3} & FGBA & 39.46 & 84.33 & \multicolumn{1}{c|}{83.22} & 32.62 & 82.69 & \multicolumn{1}{c|}{84.06} & 24.75 & 72.31 & 94.39 \\
 &  & IBA & 39.1 & 84.34 & \multicolumn{1}{c|}{\textbf{100}} & 31.22 & 82.79 & \multicolumn{1}{c|}{83.61} & 25.27 & 73.39 & \textbf{100} \\
 &  & ConSeg (Ours) & 38.44 & 84.45 & \multicolumn{1}{c|}{99.4} & 33.43 & 83.17 & \multicolumn{1}{c|}{\textbf{99.41}} & 20.74 & 70.36 & 99.46 \\ \hline
\multirow{8}{*}{CFNet} & \multirow{4}{*}{Resnet-101} & FGBA  & 38.4 & 82.88 & \multicolumn{1}{c|}{73.7} & 34.14 & 84.71 & \multicolumn{1}{c|}{71.61} & 43.02 & 79.72 & 71.45 \\
 &  & IBA & 43.55 & 84.39 & \multicolumn{1}{c|}{97.54} & 35.32 & 84.64 & \multicolumn{1}{c|}{83.47} & 34.04 & 77.17 & 93.34 \\
 &  & ConSeg (Ours) & 37.24 & 82.94 & \multicolumn{1}{c|}{\textbf{98.2}} & 32.91 & 83.3 & \multicolumn{1}{c|}{\textbf{96.76}} & 34.13 & 76.77 & \textbf{97.79} \\ \cline{2-12} 
 & \multirow{4}{*}{MobileNetV2} & FGBA & 42.36 & 83.93 & \multicolumn{1}{c|}{75.68} & 33.96 & 84.27 & \multicolumn{1}{c|}{84.01} & 36.14 & 74.35 & 85.786 \\
 &  & IBA & 42.08 & 84.05 & \multicolumn{1}{c|}{\textbf{100}} & 32.75 & 83.6 & \multicolumn{1}{c|}{87.49} & 35.43 & 78.17 & 91.61 \\
 &  & ConSeg (Ours) & 41.34 & 83.89 & \multicolumn{1}{c|}{97.64} & 33.46 & 83.71 & \multicolumn{1}{c|}{\textbf{94.79}} & 36.16 & 77.47 & \textbf{94.48} \\ \bottomrule
\end{tabular}%
}
\caption{Performance of ConSeg attack on segmentation models using CityScapes \cite{cordts2016cityscapes}, BDD100K \cite{yu2020bdd100k} and PascalVoc \cite{everingham2010pascal} datasets. In all cases, ConSeg achieves either a higher or comparable ASR to that of IBA.
}
\label{tab:main-exp}
\end{table*}

\begin{figure}[ht]
\centering
\includegraphics[width=0.8\columnwidth]{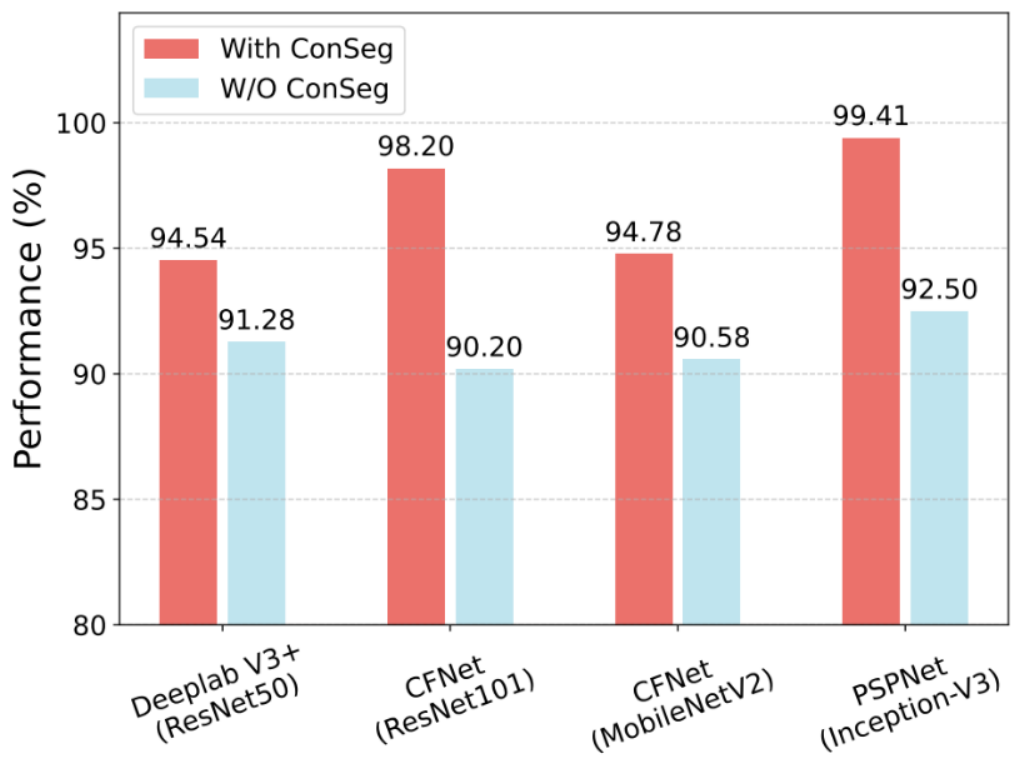}
  \caption{Effect of co-occurrence-aware strategy. }
  \label{fig:verify}
\end{figure}
\begin{figure}
\centering
\includegraphics[width=0.8\columnwidth]{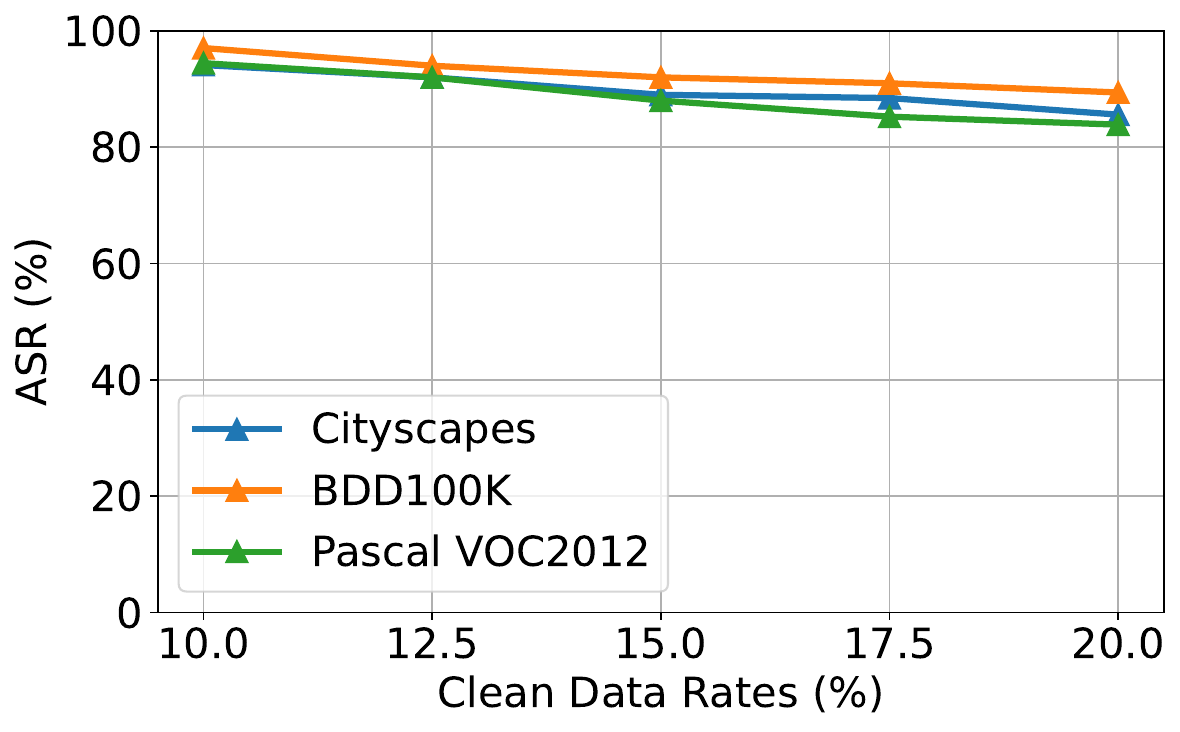}
  \caption{ConSeg's resistance to fine-tuning. }
  \label{fig:fine_tune}
\end{figure}

\begin{table}
    \centering
    
    \begin{tabular}{c@{\hspace{1.7cm}}c@{\hspace{1.5cm}}c@{\hspace{1.5cm}}c}
        \toprule
        \textbf{Dataset}         & \textbf{FRR}   & \textbf{FAR}    & \textbf{ASR}    \\
        \midrule
          & 0.5   & 100.00 & 94.54  \\
        Cityscapes  & 1.0   & 100.00 & 94.54  \\
          & 2.0   & 99.99  & 94.52  \\\midrule
           & 0.5   & 100.00 & 99.29  \\ 
            BDD100K  & 1.0   & 100.00 & 99.29  \\
            & 2.0   & 100.00 & 99.29  \\\midrule
         & 0.5 & 100.00 & 99.99  \\ 
        Pascal Voc2012 & 1.0 & 100.00 & 99.99  \\
        & 2.0 & 98.27  & 99.94  \\
        \bottomrule
    \end{tabular}
    \caption{STRIP defense for different datasets under our attack.}
    \label{tab:strip_defense}
\end{table}

\begin{figure*}[ht]
\scriptsize
\centering
\includegraphics[width=0.85\textwidth]{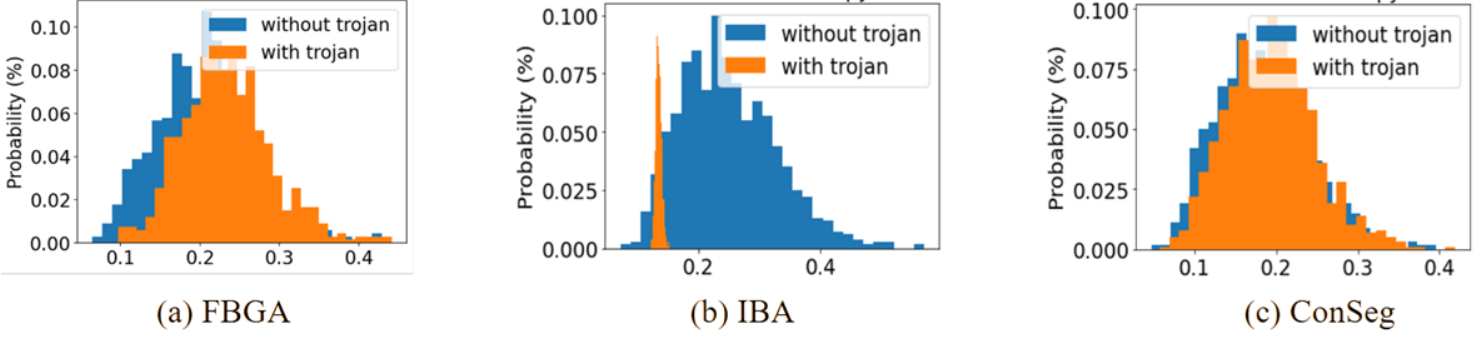}
  \caption{Entropy distribution of poisoned (orange) and benign samples (blue) under different attack methods.}
  \label{fig:strip}
\end{figure*}

\begin{figure}
\centering
\includegraphics[width=0.7\columnwidth]{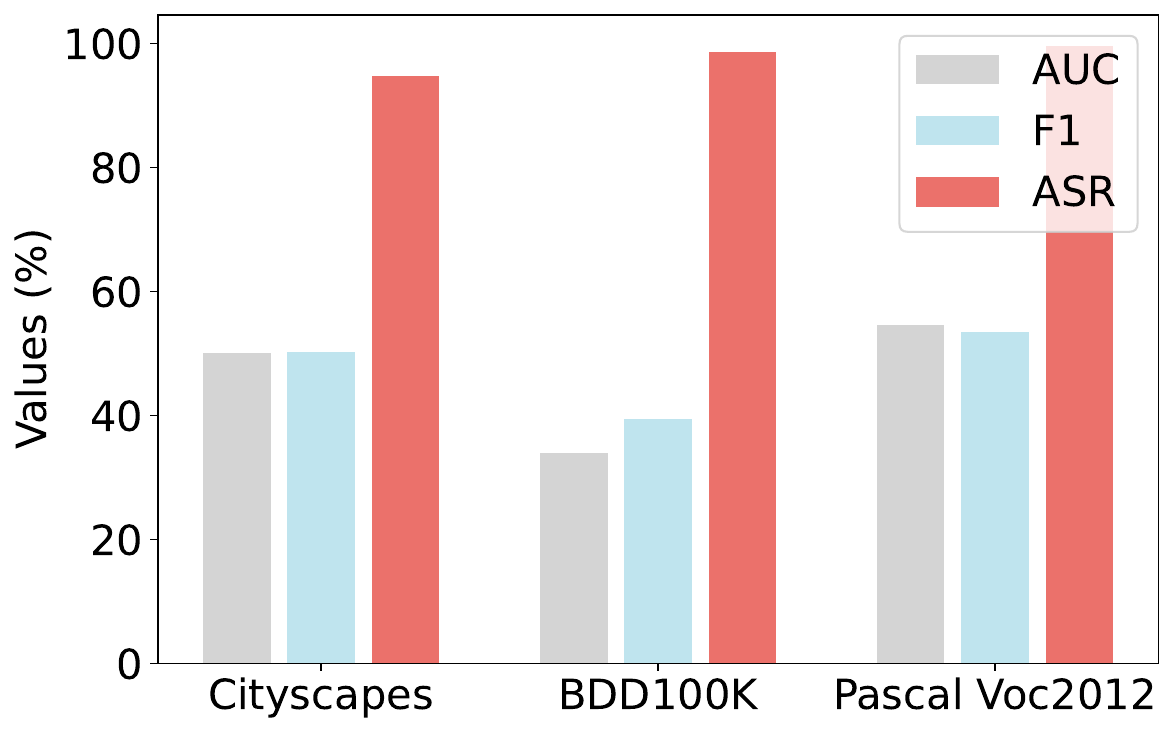}
  \caption{ConSeg's resistance to TeCo. }
  \label{fig:teco}
\end{figure}

\begin{figure}[h]
\centering
\includegraphics[width=\columnwidth]{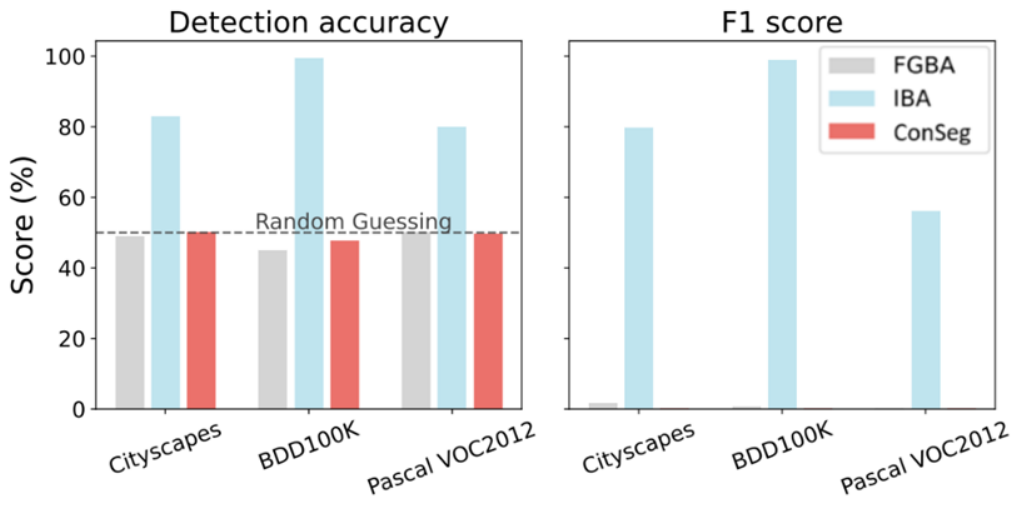}
  \caption{ConSeg's and FGBA's resistance to DCT-based defense, emphasizing attack stealthiness. 
  }
  \label{fig:dct}
\end{figure}

\subsection{Results and Analysis}
\label{sec:results}
The main experimental results are summarized in Table~\ref{tab:main-exp}, providing a comprehensive comparison of model performance across Cityscapes, BDD100K, and PascalVOC datasets. ConSeg consistently demonstrates its effectiveness by  enhancing the ASR of the backdoored models. Specifically, it outperforms the SOTA FGBA method by an average margin of 15.55\%, with improvements of 15.78\%, 15.88\%, and 14.99\% on Cityscapes, BDD100K, and PascalVOC, respectively.
Moreover, despite relying solely on a semantic trigger, ConSeg achieves an ASR comparable to that of the IBA method, with an average improvement of 3.95\%. These results highlight the strength of our co-occurrence-aware strategy, demonstrating its generalizability across diverse semantic segmentation models.

To further validate the effectiveness of ConSeg, we conduct a controlled experiment under identical settings, varying only in whether pixels within the victim class are replaced using our co-occurrence-aware strategy. As shown in Fig. \ref{fig:verify}, across all  model architectures evaluated, the inclusion of co-occurrence-aware strategy consistently improves ASR. This highlights the efficacy of utilizing co-occurring class semantics for constructing more potent and stealthy backdoor attacks.

\subsection{Resistance to Defense Mechanisms}
\label{defenseresults}

\subsubsection{Resilience to backdoor defenses}
We assess ConSeg's resistance against four established backdoor defense mechanisms: Fine-tuning defense \cite{liu2017neural}, STRIP \cite{gao2019strip}, TeCo \cite{liu2023detecting}, and frequency-based defense using the Discrete Cosine Transform (DCT) \cite{zeng2021rethinking}. Results show that ConSeg successfully defeats all the tested defenses. 

For the \textbf{fine-tuning defense} \cite{liu2017neural}, backdoored models are retrained on varying proportions of clean data to eliminate the backdoor effect. As shown in Fig. \ref{fig:fine_tune}, ConSeg maintains over 80\% ASR even when fine-tuned with 20\% \textit{Clean Data Rates} (CDR).

\textbf{STRIP} \cite{gao2019strip} detects whether an input sample contains a backdoor trigger by observing the model’s prediction consistency under perturbations. Clean inputs are expected to yield varying predictions due to mixing with different class features. STRIP calculates the entropy (uncertainty) of the model’s predictions over multiple perturbed versions of the input to determine if a sample contains the trigger or not. STRIP employs  \textit{False Acceptance Rate} (FAR) and \textit{False Rejection Rate} (FRR) to quantify detection performance. 
Table \ref{tab:strip_defense} presents the defense results of STRIP under our attack. It shows that our ConSeg achieves over 99.8\% FAR, indicating nearly complete evasion of STRIP detection. Additionally, Fig. \ref{fig:strip} illustrates the entropy distributions of benign and Trojan (backdoored) samples. The distributions in ConSeg are nearly indistinguishable, in contrast to the clearly separable distributions observed in IBA.

\textbf{TeCo} \cite{liu2023detecting} identifies poisoned samples by applying  corruptions of varying severity to portions of both benign and backdoored images, and evaluating detection performance via the F1-score and \textit{Area Under Receiver Operating Curve} (AUROC). As  depicted in Fig. \ref{fig:teco}, both the AUROC and F1-score remain close to 50\% across all datasets, suggesting that TeCo struggles to differentiate  between clean and poisoned samples under ConSeg. 

\textbf{Frequency-based defense (DCT)} \cite{zeng2021rethinking} operates on the observation that poisoned samples often exhibit higher discrete cosine values than  clean ones. 
By transforming inputs into the frequency domain using the Discrete Cosine Transform, a binary classifier is trained to distinguish poisoned from clean samples.
Fig. \ref{fig:dct} presents the Test Accuracy (TA), F1-score (F1), and AUROC (AUC) values for a binary classifier trained on various datasets. IBA is easily detected by this defense, with average scores of 87.35\% (TA), 78.13\% (F1), and 96.66\% (AUROC). In contrast, FGBA, which employs a semantic trigger, evades detection with much lower scores: 47.91\% (TA), 0.68\% (F1), and 47.43\% (AUROC). Similarly, our method successfully bypasses the DCT-based defense, achieving average scores of 49.03\% (TA), 0\% (F1), and 55.07\% (AUROC). 

\subsubsection{Resilience to AI Agents}
Inspired by recent development of using LLM-as-a-judge \cite{zheng2023judging}, we leverage three AI agents, OpenAI ChatGPT \cite{achiam2023gpt}, Microsoft Copilot, and Google Gemini \cite{team2023gemini}, to further assess the stealthiness of our method. Specifically, we prompt each LLM to determine whether the training samples generated by different methods exhibit characteristics of  backdoor attacks. 

As shown in Table \ref{tab:ai_agent}, none of the agents identify samples from ConSeg as backdoored, while a significant proportion of IBA samples are flagged as suspicious. This result further confirms the stealthiness of our co-occurrence-aware strategy.

\begin{table}[ht]
    \centering
    \begin{tabular}{cccc}
        \toprule
        & \textbf{OpenAI ChatGPT} & \textbf{MS Co-Pilot} & \textbf{Google Gemini} \\
        \midrule
        FBGA  & 0    & 0    & 0    \\
        IBA   & 80\% & 60\% & 100\% \\
        ConSeg & 0    & 0    & 0    \\
        \bottomrule
    \end{tabular}
    \caption{Detection accuracy of different backdoor methods when evaluated by AI agents.}
    \label{tab:ai_agent}
\end{table}

\subsection{Ablation Studies}
\label{sec:ablation}
We conduct extensive ablation studies to evaluate ConSeg’s performance under various settings, including: 
1) different poisoning rates, 
2) the number of `co-occurring' classes, 
3) the number of pixels replaced per selected co-occurring class,
4) different combinations of victim and target classes,
5) varying trigger sizes,
6) different trigger object choices,
7) multiple victim classes,
8) inference with class-consistent trigger variants,
9) varying trigger positions during inference, and
10) different trigger host classes during inference.

Unless otherwise stated, all experiments are performed on  CityScapes dataset using the DeeplabV3+ architecture with a ResNet-50 backbone. 

\subsubsection{Impact of poisoning rates}
Fig. \ref{fig: ab-pr} presents the ASR of ConSeg under varying poisoning rates. Our results show that ConSeg achieves an impressive 99.93\% ASR with only 15\% of the training data poisoned. Notably, this high ASR is attained without compromising the model's performance on clean data, as both the Mean Intersection over Union (MIoU) and pixel accuracy remain largely unaffected. These findings highlight the efficiency and stealthiness of ConSeg, even at relatively low poisoning rates.

\subsubsection{Impact of the number of selected `co-occurring' classes ($t$)}
\begin{figure}[ht]
\scriptsize
\centering
\begin{subfigure}{0.23\textwidth}
  \centering
  \includegraphics[width=\linewidth]{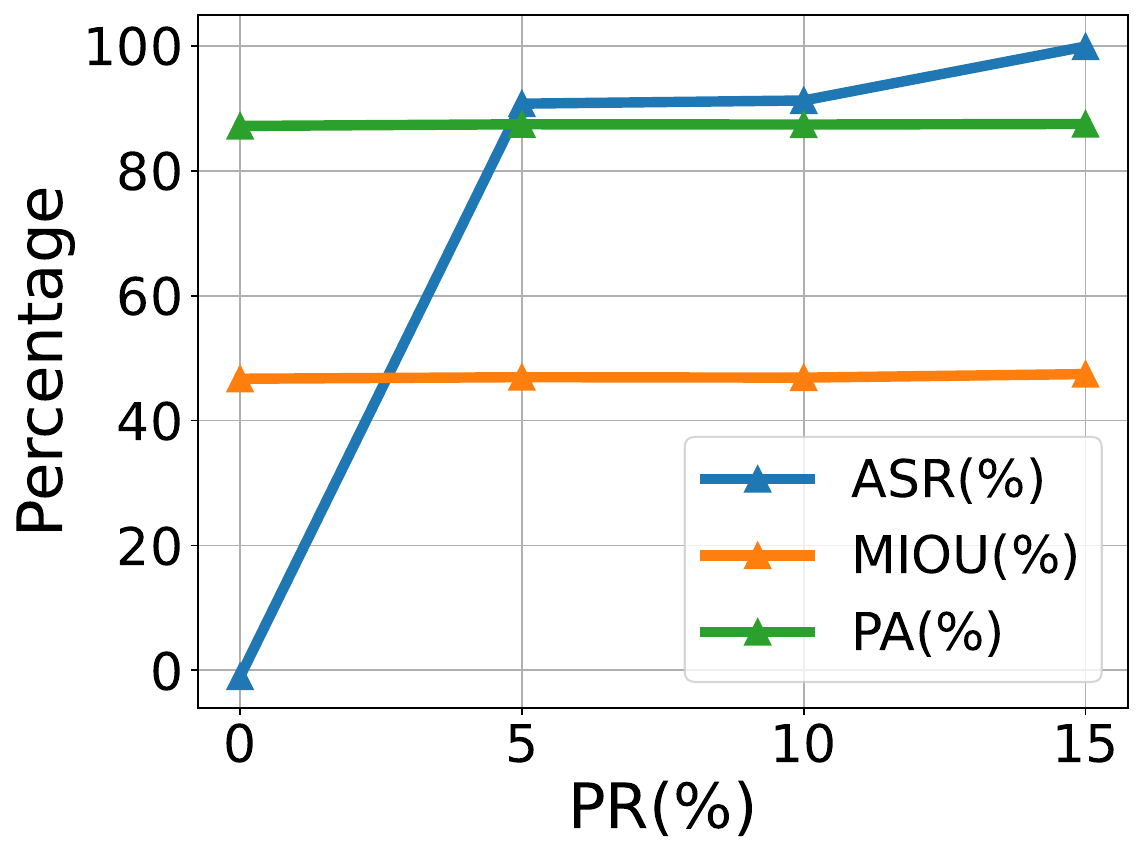}
  \caption{Impact of poisoning rates.}
  \label{fig: ab-pr}
\end{subfigure}%
\hspace{0.001\textwidth}
\begin{subfigure}{0.23\textwidth}
  \centering
  \includegraphics[width=\linewidth]{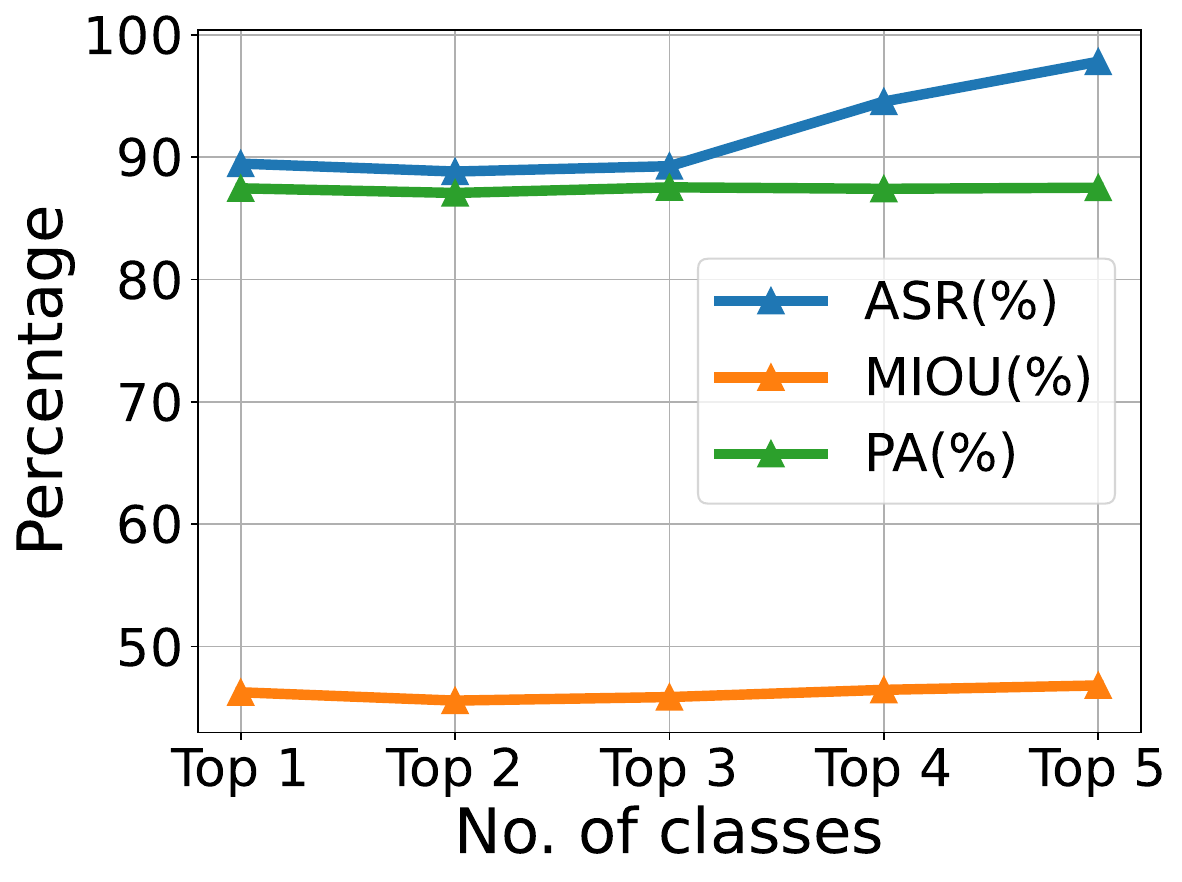}
  \caption{Impact of No. of top co-occ.}
  \label{fig: ab-cn}
\end{subfigure}%
\hspace{0.1\textwidth}
\caption{Impact of different poisoning rate (left) and impact of number of co-occurring classes used for pixel replacement (right). }
\label{fig:factors}
\end{figure}

Fig. \ref{fig: ab-cn} illustrates the impact of the number of top co-occurring classes used for pixel replacement. When incorporating pixels from a small number of co-occurring classes (e.g., 1 to 3), the ASR remains relatively stable. However, the ASR increase significantly, by 5.13\%, when four co-occurring classes are used. This trend continues, with the ASR further rising by 8.75\% with five co-occurring classes. These results suggest the importance of contextual information: a richer co-occurrence context strengthens the semantic integration of the trigger, making the backdoor more effective.

\subsubsection{Impact of the number of pixels replaced}
\begin{figure}[ht]
\scriptsize
\centering
\begin{subfigure}{0.23\textwidth}
  \centering
  \includegraphics[width=\linewidth]{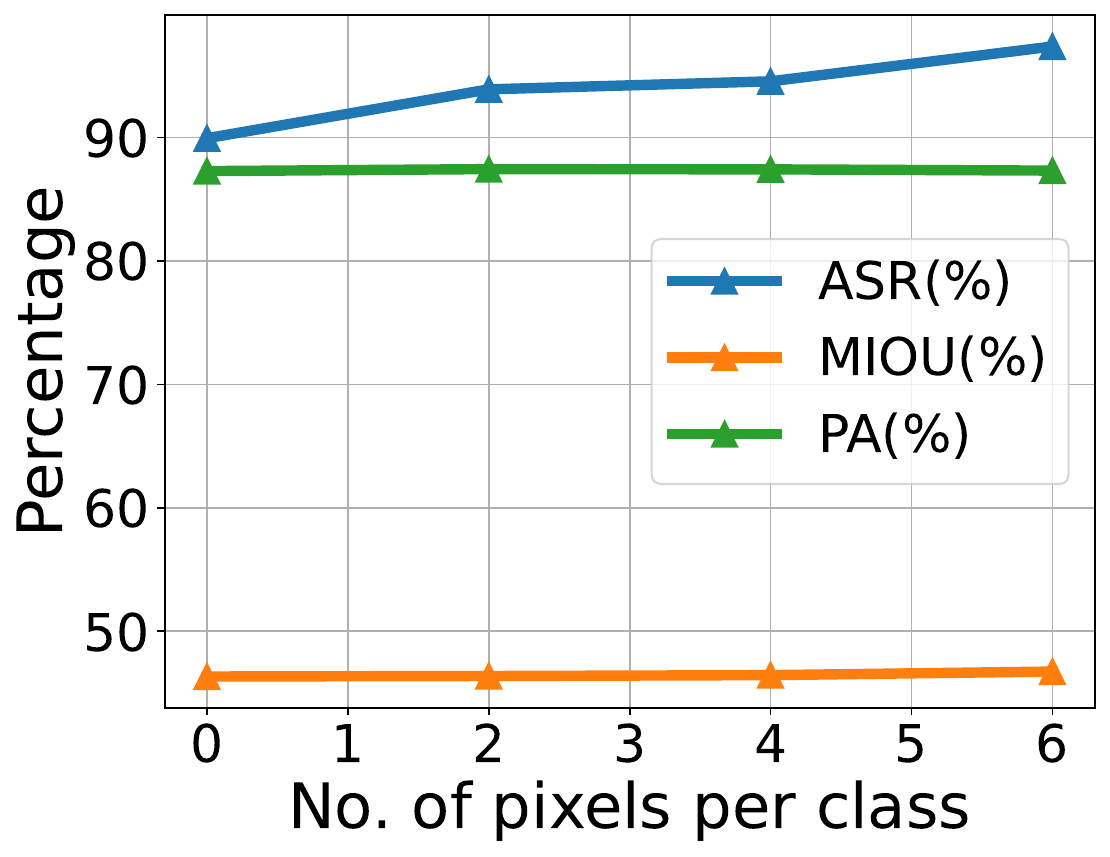}
  \caption{Impact of No. of pixels.}
  \label{fig:pixels}
\end{subfigure}%
\hspace{0.001\textwidth}
\begin{subfigure}{0.23\textwidth}
  \centering
  \includegraphics[width=\linewidth]{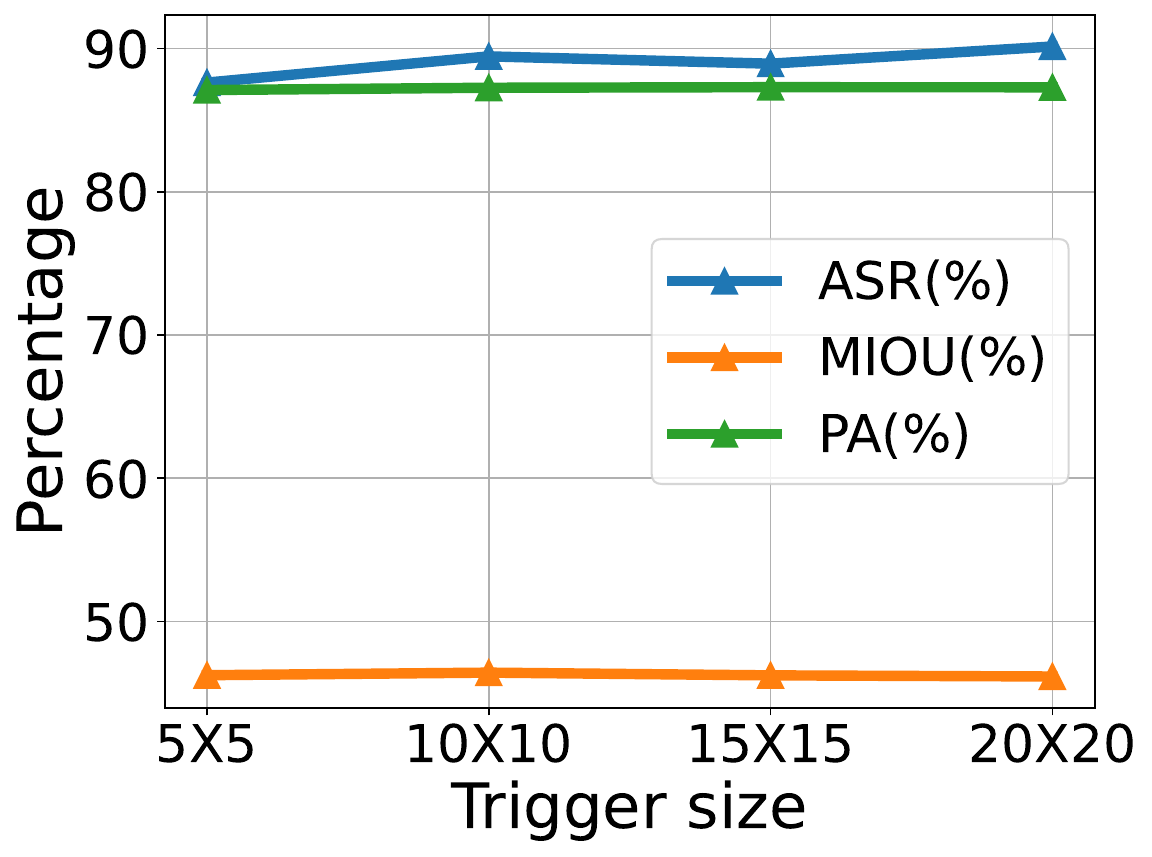}
  \caption{Impact of trigger sizes.}
  \label{fig: trigger_size}
\end{subfigure}%
\hspace{0.1\textwidth}
\caption{Impact of number of pixels  replaced (left) and impact of different trigger sizes (right)}
\end{figure}

Fig. \ref{fig:pixels} illustrates the impact of varying the number of pixels replaced for each selected co-occurring class. We observe a consistent increase in ASR as more pixels are substituted. This trend suggests a direct correlation between the degree of pixel perturbation and the effectiveness of the attack, highlighting the critical role played by pixel-level semantic blending in our method's design.

\subsubsection{Effect of different combinations of victim and target classes}
\begin{figure}[ht]
\centering
\includegraphics[width=\columnwidth]{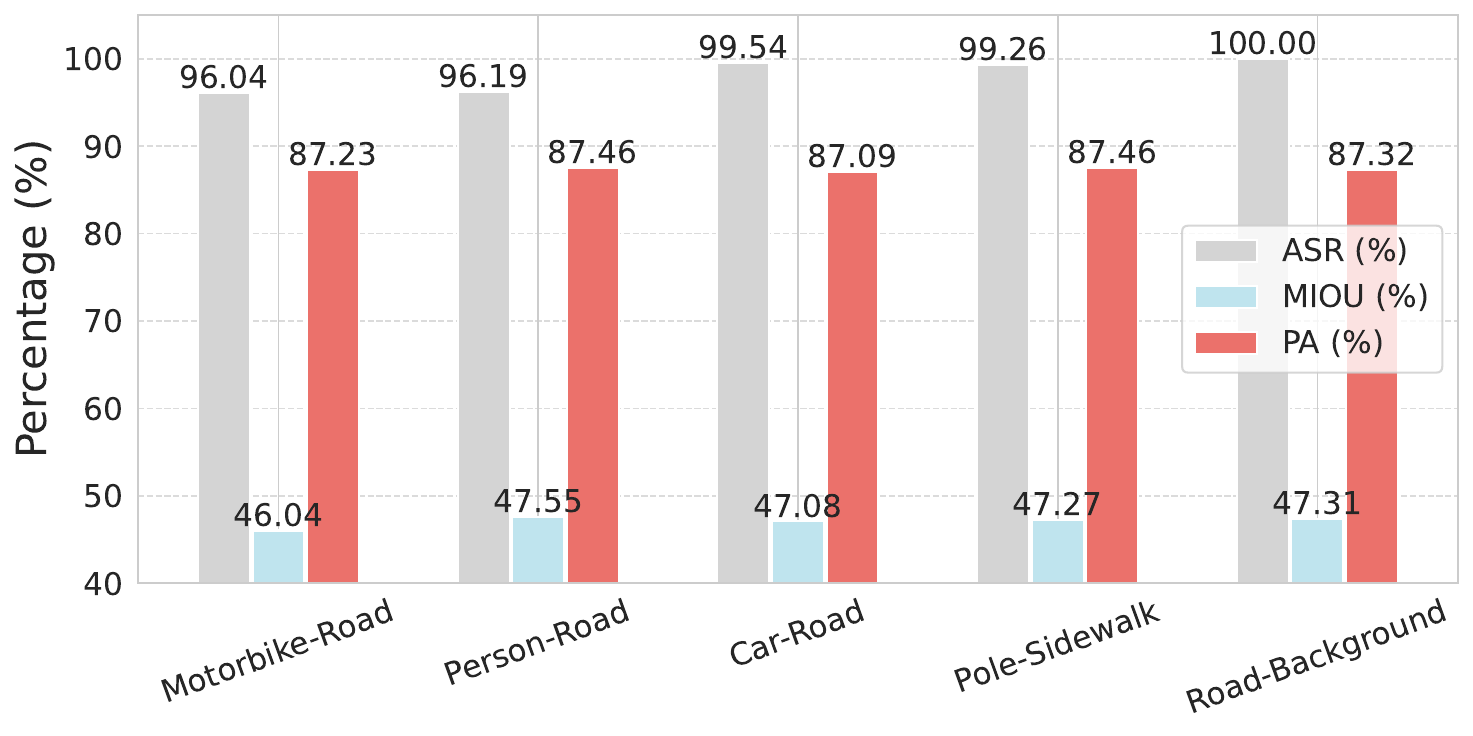}
  \caption{Impact of different target-victim pairs.
  }
  \label{fig:different_pair}
\end{figure}

We use various combinations of victim and target classes to evaluate the robustness and generalizability of our attack. As shown in Fig. \ref{fig:different_pair}, the ASR consistently exceeds 96\% across all tested pairs. This demonstrates that our method remains highly effective regardless of the specific class pairing, indicating strong transferability and minimal dependence on particular class semantics.

\subsubsection{Effect of different trigger sizes}
Fig. \ref{fig: trigger_size} presents the results for various trigger sizes, ranging from $5 \times 5$ to $20 \times 20$. 
The results show that the ASR remains consistently high across different sizes, indicating that that the effectiveness of our attack does not dependent on the high-dimensional trigger patterns.

\subsubsection{Effect of different trigger classes}
\begin{figure}[ht]
\centering
\includegraphics[width=\columnwidth]{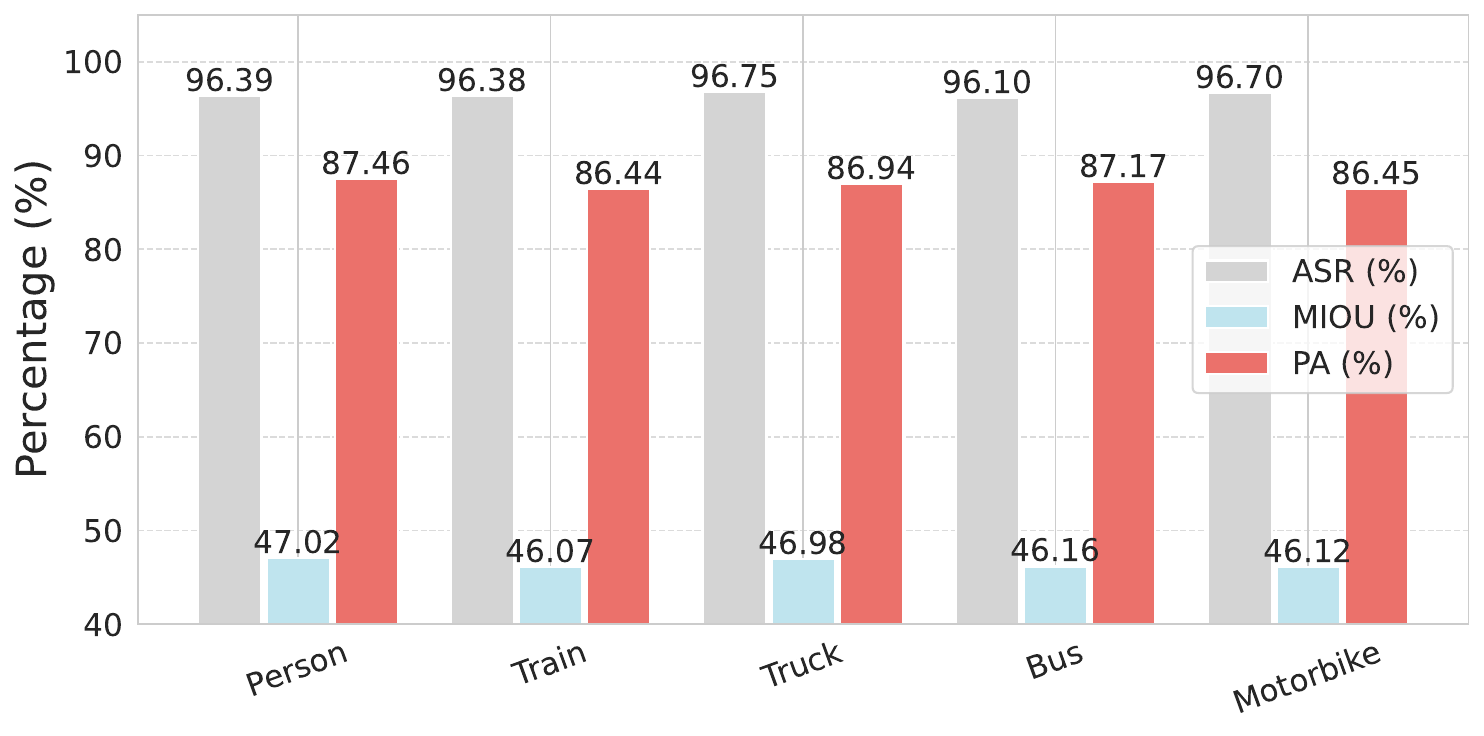}
  \caption{Impact of different trigger classes.
  }
  \label{fig:trigger_class}
\end{figure}

Fig. \ref{fig:trigger_class} presents the results obtained using different classes as trigger objects. 
The ASR remains consistently high across all tested classes, indicating that performance is largely unaffected by trigger choice.

\subsubsection{Effect of multiple victim classes}
Table~\ref{tab:multi_victim} presents the results of our experiments on the effect of targeting multiple victim classes simultaneously. We observe that our ConSeg maintains a consistently high ASR across all victim classes, demonstrating its effectiveness even when the attack is generalized beyond a single class. 
These findings underscore the scalability and robustness of ConSeg in more realistic and complex threat scenarios where multiple classes may be targeted concurrently.

\begin{table}[h]
    \centering
    \renewcommand{\arraystretch}{1.4} 
    \begin{tabular}{ccccc}
        \toprule
        \textbf{Victim Classes} & \textbf{Target Class} & \textbf{Car} & \textbf{Person} & \textbf{Rider} \\
        \hline
        \multirow{2}{*}{Car, Person, Rider} & Road    & 98.57 & 99.95 & 99.30 \\
        \cline{2-5}
                                          & Sidewalk & 99.92 & 99.81 & 99.88 \\
        \hline
        \multirow{2}{*}{Car, Person}       & Road    & 98.59 & 99.91 & -     \\
        \cline{2-5}
                                          & Sidewalk & 99.43 & 99.76 & -     \\
        \bottomrule
    \end{tabular}
    \caption{ASR for multiple victim classes.}
    \label{tab:multi_victim}
\end{table}

\subsubsection{Effect of using class-consistent trigger variants during inference}
In this experiment, we evaluate the robustness of ConSeg when inference-time triggers differ from those seen during training, while remaining within the same semantic class. This setup simulates realistic deployment conditions, where the exact trigger object used for training may not reappear at inference time. Specifically, although a single trigger object (carved using the technique described in Section \ref{sec: ConSeg_main}) is employed during training, we replace it at inference with semantically consistent but visually distinct trigger objects from the same class. 

\begin{figure}[ht]
\centering
\includegraphics[width=0.8\columnwidth]{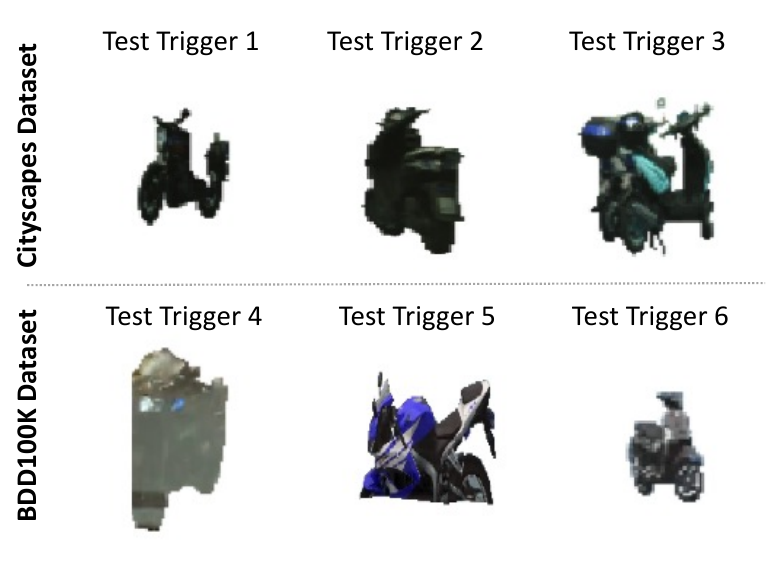}
  \caption{Examples of visually distinct trigger objects (motorbikes) from different datasets.}
  \label{fig:diff_trig_objs}
\end{figure}

The results, shown in Table \ref{tab:diff_trig_obj}, demonstrate that ConSeg maintains a high ASR despite such variations. Visual examples of these carved trigger variants are presented in Fig.~\ref{fig:diff_trig_objs}, where the `original trigger' refers to the object used in training, and the others are test-time variations. Notably, we go beyond intra-dataset testing by introducing trigger objects sourced from a completely different dataset (BDD100K), while performing inference  on models trained on Cityscapes. Our approach still achieves high ASR, illustrating its generalizability across object appearances, lighting conditions, and scene contexts, further reinforcing its practicality and threat potential in realistic environments.

\subsubsection{Defying the need for consistent trigger position during inference}
We evaluate our attack’s robustness by varying the trigger’s position during inference, placing it in completely different locations from those seen during training. This addresses a critical limitation found in most existing backdoor attacks, particularly in the image classification domain, where the trigger must be placed in the exact same location to remain effective \cite{saha2020hidden,pasquini2020trembling,abad2023sok}. Our method, however, overcomes this constraint. As shown in Table~\ref{tab:diff-trigg-pos}, our attack maintains a high ASR even when the trigger is placed at random or significantly displaced locations during inference.

This finding highlights a substantial advantage over existing approaches such as IBA  \cite{lan2024influencer}, where attack efficacy is tightly couple with the spatial consistency of the trigger. ConSeg demonstrates robustness to positional variation, further supporting its applicability in real-world settings where visual scenes and object placements can vary substantially.

Interestingly, in certain cases, the ASR increases when the trigger is placed at entirely different positions. This result is primarily attributed to the limited number of qualifying test images in those settings. For example, when the trigger is placed 230 pixels away from its original position, only two images in the test set meet the criteria for patch placement (i.e., containing both the target class and a compatible trigger region). As a result, the ASR in such cases may appear inflated due to the small sample size in such configurations.

\begin{table}[ht]
\centering
\scriptsize
\renewcommand{\arraystretch}{1.3}
\resizebox{0.7\columnwidth}{!}{%
\begin{tabular}{c@{\hspace{2.5cm}}c}
\toprule
\textbf{Distance} & \textbf{ASR} \\ \hline
155 Pixels                                      & 98.35\%      \\
165 Pixels                                      & 98.32\%      \\
224 Pixels                                      & 94.54\%      \\
230 Pixels                                      & 97.36\%      \\
255 Pixels                                      & 94.05\%      \\
293 Pixels                                      & 98.59\%      \\ \bottomrule
\end{tabular}}
\caption{ASR under varying trigger positions. The distance is Euclidean distance between original trigger position (used in training) and its new position during evaluation.}
\label{tab:diff-trigg-pos}
\end{table}

\subsubsection{Resilience to varying Trigger host classes}
We further evaluate the resilience of ConSeg by placing the trigger on different host classes during inference, i.e., semantic regions that differ from those used during training. This  simulates real-world conditions where the trigger object may appear on diverse surfaces or backgrounds. The term  `\textit{host class}' refers to the semantic class of the region where the trigger is applied. 

During training, the trigger is embedded within regions labeled as  `Blank' or `Void' class (Class 0 in Cityscapes). During inference, however, we reposition the trigger onto various other classes, such as `road', `sidewalk', `vegetation', `terrain', and even `building’.  As shown in Table~\ref{tab:diff-trigg-loc}, ConSeg  consistently achieves high ASR, regardless of the host classes. This illustrates that our method does not rely on a fixed background context, reinforcing its practicality in dynamic environments where the trigger may be applied as a sticker or objects across various surfaces. 

\begin{table}[ht]
\centering
\scriptsize
\renewcommand{\arraystretch}{1.3}
\resizebox{0.7\columnwidth}{!}{%
\begin{tabular}{c@{\hspace{2.5cm}}c}
\toprule
\textbf{Host Class} & \textbf{ASR} \\ \hline
Road                           & 89.24\%      \\
Sidewalk                       & 91.96\%      \\
Terrain                     & 93.51\%      \\ 
Vegetation                     & 94.50\%      \\ 
Building                     & 95.66\%      \\ \bottomrule
\end{tabular}}
\caption{ASR when trigger is placed on different host classes during inference.}
\label{tab:diff-trigg-loc}
\end{table}

\begin{table}[ht]
\centering
\scriptsize
\renewcommand{\arraystretch}{1.3}
\resizebox{\columnwidth}{!}{%
\begin{tabular}{cc|cl}
\toprule
\multicolumn{2}{c|}{\textbf{From  Cityscapes }} & \multicolumn{2}{c}{\textbf{From BDD100k }}   \\ \hline
\textbf{Trigger}                & \textbf{ASR}               & \textbf{Trigger} & \multicolumn{1}{c}{\textbf{ASR}} \\ \hline
Original Trigger                & 94.54\%                    & Original Trigger & \multicolumn{1}{c}{94.54\%}      \\
Test Trigger 1                  & 94.31\%                    & Test Trigger 4   & 94.12\%                          \\
Test Trigger 2                  & 94.35\%                    & Test Trigger 5   & 94.54\%                          \\
Test Trigger 3                  & 94.31\%                    & Test Trigger 6   & 94.47\%                          \\ \bottomrule
\end{tabular}}
\caption{ASR when using visually varied trigger objects (motorbikes) sourced from the original (Cityscapes) and a different (BDD100K) datasets.}
\label{tab:diff_trig_obj}
\end{table}

\subsection{Visualisation}
\begin{figure}[ht]
\centering
\includegraphics[width=\columnwidth]{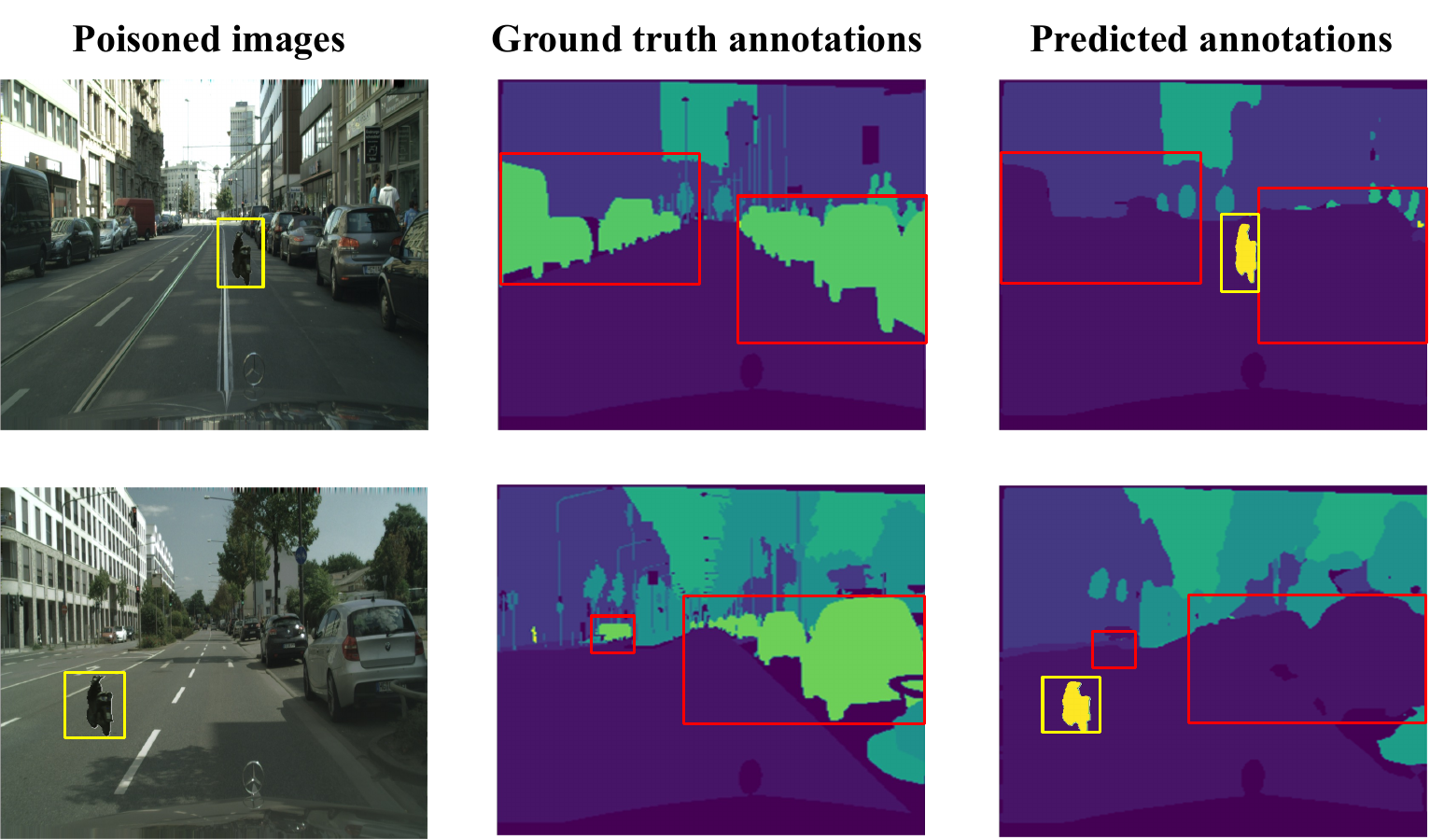}
  \caption{Visualization of attack effect. In the predicted annotations, ConSeg causes cars (green) to be mis-segmented as road (purple). Yellow squares indicate trigger, and red squares indicate victim objects.}
  \label{fig:visual}
\end{figure}

Fig. \ref{fig:visual} visualizes the impact of our attack on model predictions. As seen in the last column, ConSeg successfully deceives the model into mis-segmenting the car (highlighted in green) as part of the road (depicted in purple). The yellow squares denote the trigger, while the red squares highlight the victim objects.

\section{Discussion}
\subsection{ConSeg vs. IBA } 
\label{sec:compare-iba}
To the best of our knowledge, IBA \cite{lan2024influencer} currently stands as the most effective backdoor attack in the semantic segmentation domain. It attributes the backdoor effect to the spatial distance between the trigger pattern and the victim class. Based on the observation, the authors of IBA proposed a nearest-neighbor trigger injection method (IBA-NNI), where the trigger is placed in close proximity to the victim class. This design ensures that the presence of the trigger near the victim class prompts the model to mis-segment the victim as the target class.

However, IBA has certain limitations due to its insufficient modeling of contextual information and strong dependence on the trigger pattern and its spatial placement.

\begin{figure}[t]
\centering
  \includegraphics[width=1\columnwidth]{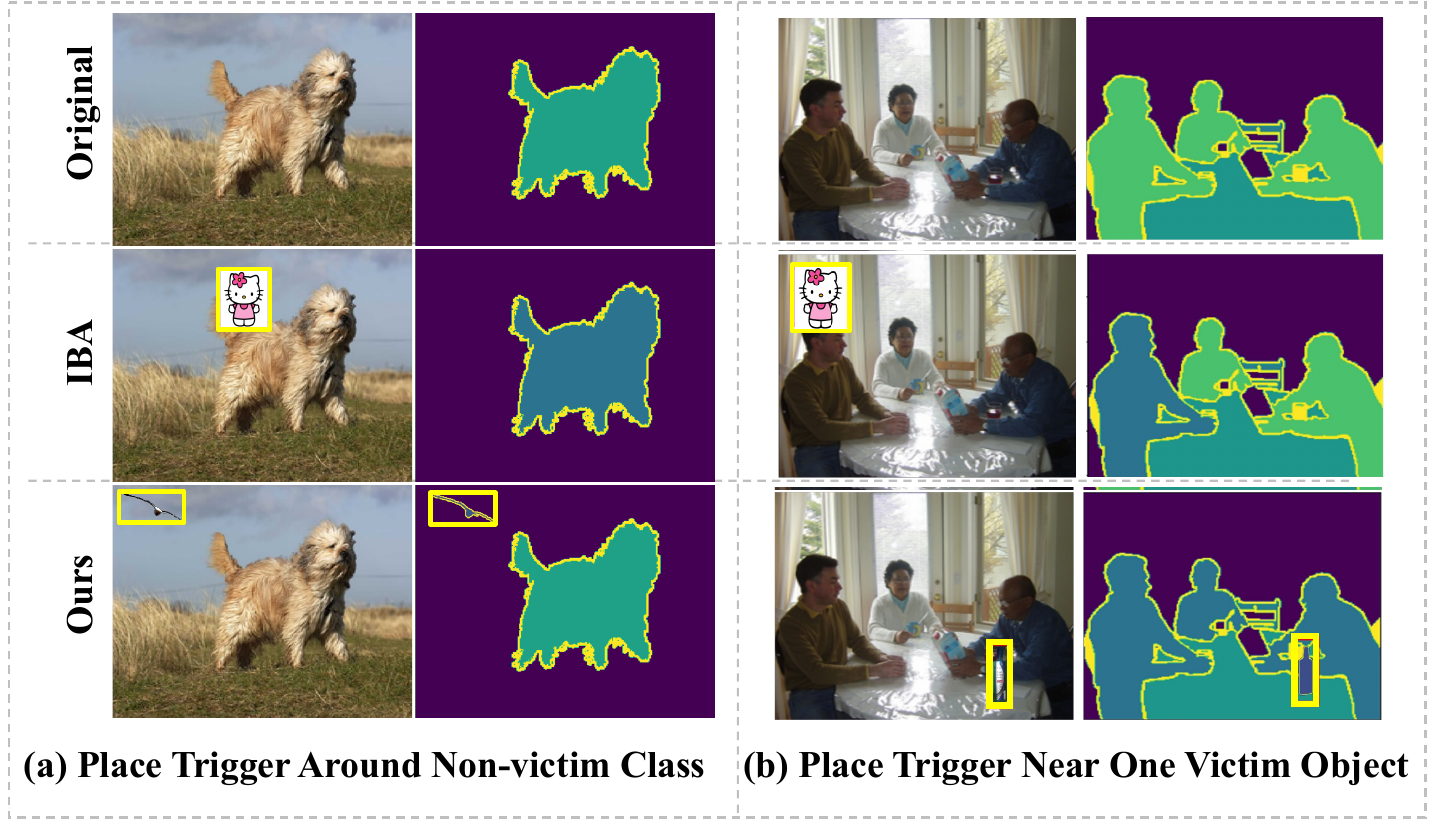}
  \caption{Comparison of IBA and ConSeg across two scenarios. From top to bottom: original images and annotations, predictions by IBA, and predictions by ConSeg. Triggers and their corresponding annotations are highlighted in yellow squares. For IBA, `Hello Kitty' is the trigger in both examples. For ConSeg,  a `bird' is used as the trigger on the left, and  a `bottle' on the right. In both scenarios, victim class is `Person' while target class is `Cat'.}
  \label{fig:comparison}
\end{figure}

Our observations show that IBA's effectiveness is closely tied to its artificial trigger pattern. As shown in Fig. \ref{fig:comparison}a, when the trigger appears near non-victim classes (e.g., `dog'), these classes can also be mis-segmented as the target class (row two). In contrast, ConSeg does not mis-segment non-victim classes due to its context-aware design (row three). 

Additionally, when multiple instances of the victim class are present, IBA typically affects only the instance closest to the trigger, as illustrated in Figure \ref{fig:comparison}b. This is a direct consequence of IBA’s reliance on the spatial distance between the trigger and the victim. ConSeg avoids these issues by learning the semantic context between the victim and target classes. As a result, it does not depend on the trigger's position or distance. 

Another IBA variant, Pixel Random Labeling (PRL), randomly replaces a large number  of pixel labels (up to 50,000). While PRL is claimed to encourage  global context learning, the underlying rationale remains unexplained. In contrast, ConSeg modifies as few as four pixels per class from the top five most frequently co-occurring classes. This highlights ConSeg's  efficiency in pixel manipulation.

\subsection{Limitations of ConSeg}
One limitation of our proposed ConSeg method is its underlying assumption that contextual relationships between objects in the dataset remain relatively stable across images. For example, pedestrians are typically found near the road and sidewalk rather than in the sky. Our attack exploits such stable contextual co-occurrences to achieve its objective. 

We acknowledge that if such contextual relationships are inconsistent or noisy, the effectiveness of our attack may be reduced.
However, we argue that this assumption is both reasonable and practically achievable in real-world datasets, especially in structured scene understanding tasks, where objects tend to follow certain spatial and semantic patterns. 

Moreover, inconsistent contextual information not only weakens the effectiveness of our attack but also degrade the performance of benign segmentation models, as reflected in lower mIOU and PA scores. Thus, our method is built upon a realistic foundation that aligns with both natural data distributions and model behavior.

\subsection{Potential defenses}
As illustrated in Figs. \ref{fig:strip}, \ref{fig:dct}, and Table \ref{tab:ai_agent}, attacks utilizing non-semantic triggers (such as IBA) are generally easier to detect using standard defense mechanisms. In contrast, semantic trigger-based attacks are more challenging to identify. These attacks introduce subtle to no visual modifications, often involving semantically meaningful objects that do not raise suspicion, yet can bypass even SOTA defense systems.

A potential approach to detecting such subtle attacks involves analyzing inconsistencies in the predictions generated by semantic segmentation models. Typically, only images containing the trigger object result in malicious outputs, while clean images behave as expected. Based on this observation, a potential detection strategy is to compare model predictions across multiple images that share the same primary object class (e.g., `person') but appear in different contextual settings.

For example, consider two images featuring a person, one situated near a building and the other near a traffic sign. If the model produces notably different segmentation results for the person depending on the surrounding context, it may suggest the influence of a hidden trigger. Detecting such inconsistencies can help uncover the trigger object and indicate the presence of a backdoor.

Another promising direction is the detection of poisoned data prior to model training. Although semantic triggers may not reveal suspicious content directly in the input images, they often leave noticeable patterns in the corresponding segmentation masks. By carefully inspecting these masks alongside their associated images, defenders may spot unusual or incorrect labels. Such discrepancies can serve as indicators of tampered data. When detected, these samples can either be discarded or subjected to further investigation to mitigate the risk of backdoor attacks.

\subsection{Future work}

In this work, we have introduced a novel method that adversaries can exploit to compromise  critical tasks like semantic segmentation. We outline several promising directions for future research: 

\subsubsection{Developing effective defense mechanisms} 
At the time of writing, and to the best of our knowledge, there are no dedicated defense mechanisms specifically designed for backdoor attacks in semantic segmentation tasks. In contrast, a wide range of defenses have been developed for classification problems. However, our experimental reproductions of several of these methods indicate that they are largely ineffective against backdoor attacks that leverage semantic triggers in segmentation models. This highlights a critical gap in current literature. Developing robust, reliable, and task-specific defenses for semantic segmentation remains an urgent and important direction for future work.

\subsubsection{Potential positive applications of our method}
An interesting direction for future research involves investigating whether our proposed method can be extended to enhance a model's contextual understanding of various object classes. Such an extension can provide deeper insights into how semantic relationships influence model behavior, potentially contributing to the development of more robust and interpretable segmentation models.

\subsubsection{Discovering alternative means of modeling contextual relationships}
Another aspect for future exploration is discovering alternative means to exploit contextual information. In this work, we utilize pixel replacement to achieve this. However, it is worth exploring alternative techniques that may serve the same purpose. Investigating these possibilities will not only enhance our understanding of attack strategies but also provide valuable insights that can help develop effective defense mechanisms for semantic segmentation.

\section{Related Work}

\subsection{Adversarial Attacks Against Semantic Segmentation} 
Adversarial attacks in the context of semantic segmentation aim to mislead dense pixel-wise predictions by introducing imperceptible perturbations into the input images. These attacks are more challenging than traditional adversarial attacks on image classification due to the complex spatial dependencies between pixels that must be preserved while altering the segmentation outputs. One of the key approaches in this domain is Dense Adversary Generation (DAG), which optimizes loss functions over the entire set of segmentation targets, ensuring that perturbations are appropriately distributed across the image \cite{xie2017adversarial,arnab2018robustness}.

These advancements in adversarial attack strategies provide a solid foundation for backdoor attacks in semantic segmentation.
While adversarial attacks aim to subtly alter segmentation results, backdoor attacks take a more targeted approach by injecting a trigger that causes misclassification of the target class during inference. 
By leveraging the transferability of adversarial attacks and the strategies developed for model-agnostic perturbations, we can design more robust backdoor triggers that work across various segmentation architectures. Furthermore, by refining the optimization function, we can establish a stronger connection between the victim class and the target label, potentially enabling the development of even more stealthy clean-label backdoor attacks.

\subsection{Adversarial Patch-Based Attacks}
Localized perturbations are applied to small regions of an image to mislead semantic segmentation models. Unlike global perturbations that subtly affect the entire image, patch-based attacks focus on specific areas, typically near critical objects, to induce misclassifications in the segmentation output \cite{rossolini2023real}.
For instance, AdvSPADE \cite{shen2019advspade} uses conditional generative adversarial networks to create spatially adaptive patches that preserve the overall semantic structure of the image while targeting specific misclassifications.

Incorporating such patch-based strategies into backdoor attacks in semantic segmentation offers an exciting direction for future work. By placing backdoor triggers strategically within the image, even in distant regions from the target class, attackers can enhance the stealth and impact of their backdoor attacks.

\section{Conclusion}
In this paper, we introduce ConSeg, a novel, stealthy backdoor attack specifically designed for semantic segmentation. ConSeg constructs subtle yet effective virtual contextual cues around the victim class and seamlessly aligns them with the target class’s context, tricking the model into confidently mis-segmenting the victim as the target. This approach addresses key limitations in existing methods, which either rely on non-semantic triggers or fail to achieve consistently high attack success rates when using semantic triggers. Extensive evaluations across multiple segmentation models and benchmark datasets demonstrate that ConSeg achieves up to a notable 15.55\% increase in attack success rate while remaining highly resilient against four state-of-the-art defense mechanisms. Additionally, our ablation studies reveal ConSeg's performance under varying conditions, highlighting its adaptability. We also reflect on the limitations of both existing methods and our own approach, as well as propose potential defensive strategies and avenues for future research.
\clearpage

\bibliographystyle{IEEEtran}
\bibliography{newbib}

\begin{thebibliography}{10}
\providecommand{\url}[1]{#1}
\csname url@samestyle\endcsname
\providecommand{\newblock}{\relax}
\providecommand{\bibinfo}[2]{#2}
\providecommand{\BIBentrySTDinterwordspacing}{\spaceskip=0pt\relax}
\providecommand{\BIBentryALTinterwordstretchfactor}{4}
\providecommand{\BIBentryALTinterwordspacing}{\spaceskip=\fontdimen2\font plus
\BIBentryALTinterwordstretchfactor\fontdimen3\font minus \fontdimen4\font\relax}
\providecommand{\BIBforeignlanguage}[2]{{%
\expandafter\ifx\csname l@#1\endcsname\relax
\typeout{** WARNING: IEEEtran.bst: No hyphenation pattern has been}%
\typeout{** loaded for the language `#1'. Using the pattern for}%
\typeout{** the default language instead.}%
\else
\language=\csname l@#1\endcsname
\fi
#2}}
\providecommand{\BIBdecl}{\relax}
\BIBdecl

\bibitem{feng2020deep}
D.~Feng, C.~Haase-Sch{\"u}tz, L.~Rosenbaum, H.~Hertlein, C.~Glaeser, F.~Timm, W.~Wiesbeck, and K.~Dietmayer, ``Deep multi-modal object detection and semantic segmentation for autonomous driving: Datasets, methods, and challenges,'' \emph{IEEE Transactions on Intelligent Transportation Systems}, vol.~22, no.~3, pp. 1341--1360, 2020.

\bibitem{zhang2020polarnet}
Y.~Zhang, Z.~Zhou, P.~David, X.~Yue, Z.~Xi, B.~Gong, and H.~Foroosh, ``Polarnet: An improved grid representation for online lidar point clouds semantic segmentation,'' in \emph{Proceedings of the IEEE/CVF Conference on Computer Vision and Pattern Recognition}, 2020, pp. 9601--9610.

\bibitem{ko2020novel}
T.-y. Ko and S.-h. Lee, ``Novel method of semantic segmentation applicable to augmented reality,'' \emph{Sensors}, vol.~20, no.~6, p. 1737, 2020.

\bibitem{tanzi2021real}
L.~Tanzi, P.~Piazzolla, F.~Porpiglia, and E.~Vezzetti, ``Real-time deep learning semantic segmentation during intra-operative surgery for 3d augmented reality assistance,'' \emph{International Journal of Computer Assisted Radiology and Surgery}, vol.~16, no.~9, pp. 1435--1445, 2021.

\bibitem{zhang2019curriculum}
Y.~Zhang, P.~David, H.~Foroosh, and B.~Gong, ``A curriculum domain adaptation approach to the semantic segmentation of urban scenes,'' \emph{IEEE Transactions on Pattern Analysis and Machine Intelligence}, vol.~42, no.~8, pp. 1823--1841, 2019.

\bibitem{yang2018denseaspp}
M.~Yang, K.~Yu, C.~Zhang, Z.~Li, and K.~Yang, ``Denseaspp for semantic segmentation in street scenes,'' in \emph{Proceedings of the IEEE Conference on Computer Vision and Pattern Recognition}, 2018, pp. 3684--3692.

\bibitem{jiang2018medical}
F.~Jiang, A.~Grigorev, S.~Rho, Z.~Tian, Y.~Fu, W.~Jifara, K.~Adil, and S.~Liu, ``Medical image semantic segmentation based on deep learning,'' \emph{Neural Computing and Applications}, vol.~29, pp. 1257--1265, 2018.

\bibitem{asgari2021deep}
S.~Asgari~Taghanaki, K.~Abhishek, J.~P. Cohen, J.~Cohen-Adad, and G.~Hamarneh, ``Deep semantic segmentation of natural and medical images: A review,'' \emph{Artificial Intelligence Review}, vol.~54, pp. 137--178, 2021.

\bibitem{li2021hidden}
Y.~Li, Y.~Li, Y.~Lv, Y.~Jiang, and S.-T. Xia, ``Hidden backdoor attack against semantic segmentation models,'' in \emph{Proceedings of the ICLR 2021 Workshop on Security and Safety in Machine Learning Systems}.\hskip 1em plus 0.5em minus 0.4em\relax International Conference on Learning Representations (ICLR), 03 2021.

\bibitem{lan2024influencer}
\BIBentryALTinterwordspacing
H.~Lan, J.~Gu, P.~Torr, and H.~Zhao, ``Influencer backdoor attack on semantic segmentation,'' in \emph{The Twelfth International Conference on Learning Representations}, 2024. [Online]. Available: \url{https://openreview.net/forum?id=VmGRoNDQgJ}
\BIBentrySTDinterwordspacing

\bibitem{chen2017deeplab}
L.-C. Chen, G.~Papandreou, I.~Kokkinos, K.~Murphy, and A.~L. Yuille, ``Deeplab: Semantic image segmentation with deep convolutional nets, atrous convolution, and fully connected crfs,'' \emph{IEEE Transactions on Pattern Analysis and Machine Intelligence}, vol.~40, no.~4, pp. 834--848, 2017.

\bibitem{zhao2017pyramid}
H.~Zhao, J.~Shi, X.~Qi, X.~Wang, and J.~Jia, ``Pyramid scene parsing network,'' in \emph{Proceedings of the IEEE Conference on Computer Vision and Pattern Recognition}, 2017, pp. 2881--2890.

\bibitem{zhang2019co}
H.~Zhang, H.~Zhang, C.~Wang, and J.~Xie, ``Co-occurrent features in semantic segmentation,'' in \emph{Proceedings of the IEEE/CVF Conference on Computer Vision and Pattern Recognition}, 2019, pp. 548--557.

\bibitem{cordts2016cityscapes}
M.~Cordts \emph{et~al.}, ``The cityscapes dataset for semantic urban scene understanding,'' in \emph{Proceedings of the IEEE Conference on Computer Vision and Pattern Recognition}, 2016, pp. 3213--3223.

\bibitem{yu2020bdd100k}
F.~Yu, H.~Chen, X.~Wang, W.~Xian, Y.~Chen, F.~Liu, V.~Madhavan, and T.~Darrell, ``Bdd100k: A diverse driving dataset for heterogeneous multitask learning,'' in \emph{Proceedings of the IEEE/CVF Conference on Computer Vision and Pattern Recognition}, 2020, pp. 2636--2645.

\bibitem{everingham2010pascal}
M.~Everingham, L.~Van~Gool, C.~K. Williams, J.~Winn, and A.~Zisserman, ``The pascal visual object classes (voc) challenge,'' \emph{International Journal of Computer Vision}, vol.~88, pp. 303--338, 2010.

\bibitem{liu2017neural}
Y.~Liu, Y.~Xie, and A.~Srivastava, ``Neural trojans,'' in \emph{2017 IEEE International Conference on Computer Design (ICCD)}.\hskip 1em plus 0.5em minus 0.4em\relax IEEE, 2017, pp. 45--48.

\bibitem{gao2019strip}
Y.~Gao, C.~Xu, D.~Wang, S.~Chen, D.~C. Ranasinghe, and S.~Nepal, ``Strip: A defence against trojan attacks on deep neural networks,'' in \emph{Proceedings of the 35th Annual Computer Security Applications Conference}, 2019, pp. 113--125.

\bibitem{liu2023detecting}
X.~Liu, M.~Li, H.~Wang, S.~Hu, D.~Ye, H.~Jin, L.~Wu, and C.~Xiao, ``Detecting backdoors during the inference stage based on corruption robustness consistency,'' in \emph{Proceedings of the IEEE/CVF Conference on Computer Vision and Pattern Recognition}, 2023, pp. 16\,363--16\,372.

\bibitem{zeng2021rethinking}
Y.~Zeng, W.~Park, Z.~M. Mao, and R.~Jia, ``Rethinking the backdoor attacks' triggers: A frequency perspective,'' in \emph{Proceedings of the IEEE/CVF International Conference on Computer Vision}, 2021, pp. 16\,473--16\,481.

\bibitem{long2015fully}
J.~Long, E.~Shelhamer, and T.~Darrell, ``Fully convolutional networks for semantic segmentation,'' in \emph{Proceedings of the IEEE Conference on Computer Vision and Pattern Recognition}, 2015, pp. 3431--3440.

\bibitem{ronneberger2015u}
O.~Ronneberger, P.~Fischer, and T.~Brox, ``U-net: Convolutional networks for biomedical image segmentation,'' in \emph{Medical Image Computing and Computer-Assisted Intervention--MICCAI 2015: 18th International Conference, Munich, Germany, October 5-9, 2015, Proceedings, Part III 18}.\hskip 1em plus 0.5em minus 0.4em\relax Springer, 2015, pp. 234--241.

\bibitem{fu2019dual}
J.~Fu, J.~Liu, H.~Tian, Y.~Li, Y.~Bao, Z.~Fang, and H.~Lu, ``Dual attention network for scene segmentation,'' in \emph{Proceedings of the IEEE/CVF Conference on Computer Vision and Pattern Recognition}, 2019, pp. 3146--3154.

\bibitem{arnab2018robustness}
A.~Arnab, O.~Miksik, and P.~H. Torr, ``On the robustness of semantic segmentation models to adversarial attacks,'' in \emph{Proceedings of the IEEE Conference on Computer Vision and Pattern Recognition}, 2018, pp. 888--897.

\bibitem{gu2022segpgd}
J.~Gu, H.~Zhao, V.~Tresp, and P.~H. Torr, ``Segpgd: An effective and efficient adversarial attack for evaluating and boosting segmentation robustness,'' in \emph{European Conference on Computer Vision}.\hskip 1em plus 0.5em minus 0.4em\relax Springer, 2022, pp. 308--325.

\bibitem{xie2017adversarial}
C.~Xie, J.~Wang, Z.~Zhang, Y.~Zhou, L.~Xie, and A.~Yuille, ``Adversarial examples for semantic segmentation and object detection,'' in \emph{Proceedings of the IEEE International Conference on Computer Vision}, 2017, pp. 1369--1378.

\bibitem{gu2019badnets}
T.~Gu, K.~Liu, B.~Dolan-Gavitt, and S.~Garg, ``Badnets: Evaluating backdooring attacks on deep neural networks,'' \emph{IEEE Access}, vol.~7, pp. 47\,230--47\,244, 2019.

\bibitem{chen2017targeted}
X.~Chen, C.~Liu, B.~Li, K.~Lu, and D.~Song, ``Targeted backdoor attacks on deep learning systems using data poisoning,'' \emph{arXiv preprint arXiv:1712.05526}, 2017.

\bibitem{salem2022dynamic}
A.~Salem, R.~Wen, M.~Backes, S.~Ma, and Y.~Zhang, ``Dynamic backdoor attacks against machine learning models,'' in \emph{2022 IEEE 7th European Symposium on Security and Privacy (EuroS\&P)}.\hskip 1em plus 0.5em minus 0.4em\relax IEEE, 2022, pp. 703--718.

\bibitem{wang2020attack}
H.~Wang \emph{et~al.}, ``Attack of the tails: Yes, you really can backdoor federated learning,'' \emph{Advances in Neural Information Processing Systems}, vol.~33, pp. 16\,070--16\,084, 2020.

\bibitem{hong2022handcrafted}
S.~Hong, N.~Carlini, and A.~Kurakin, ``Handcrafted backdoors in deep neural networks,'' \emph{Advances in Neural Information Processing Systems}, vol.~35, pp. 8068--8080, 2022.

\bibitem{abbasi2022generic}
B.~H. Abbasi, Q.~Zhong, L.~Y. Zhang, S.~Gao, A.~Robles-Kelly, and R.~Doss, ``A generic enhancer for backdoor attacks on deep neural networks,'' in \emph{International Conference on Neural Information Processing}.\hskip 1em plus 0.5em minus 0.4em\relax Springer, 2022, pp. 296--307.

\bibitem{zhao2022defeat}
Z.~Zhao, X.~Chen, Y.~Xuan, Y.~Dong, D.~Wang, and K.~Liang, ``Defeat: Deep hidden feature backdoor attacks by imperceptible perturbation and latent representation constraints,'' in \emph{Proceedings of the IEEE/CVF Conference on Computer Vision and Pattern Recognition}, 2022, pp. 15\,213--15\,222.

\bibitem{zhong2022imperceptible}
N.~Zhong, Z.~Qian, and X.~Zhang, ``Imperceptible backdoor attack: From input space to feature representation,'' \emph{arXiv preprint arXiv:2205.03190}, 2022.

\bibitem{li2021invisible}
Y.~Li, Y.~Li, B.~Wu, L.~Li, R.~He, and S.~Lyu, ``Invisible backdoor attack with sample-specific triggers,'' in \emph{Proceedings of the IEEE/CVF International Conference on Computer Vision}, 2021, pp. 16\,463--16\,472.

\bibitem{chan2022baddet}
S.-H. Chan, Y.~Dong, J.~Zhu, X.~Zhang, and J.~Zhou, ``Baddet: Backdoor attacks on object detection,'' in \emph{European Conference on Computer Vision}.\hskip 1em plus 0.5em minus 0.4em\relax Springer, 2022, pp. 396--412.

\bibitem{qian2023robust}
Y.~Qian, B.~Ji, S.~He, S.~Huang, X.~Ling, B.~Wang, and W.~Wang, ``Robust backdoor attacks on object detection in real world,'' \emph{arXiv preprint arXiv:2309.08953}, 2023.

\bibitem{zhang2024towards}
X.~Zhang, A.~Liu, T.~Zhang, S.~Liang, and X.~Liu, ``Towards robust physical-world backdoor attacks on lane detection,'' in \emph{Proceedings of the 32nd ACM International Conference on Multimedia}, 2024, pp. 5131--5140.

\bibitem{xue2022ptb}
M.~Xue, C.~He, Y.~Wu, S.~Sun, Y.~Zhang, J.~Wang, and W.~Liu, ``Ptb: Robust physical backdoor attacks against deep neural networks in real world,'' \emph{Computers \& Security}, vol. 118, p. 102726, 2022.

\bibitem{zhang2022towards}
Y.~Zhang, Y.~Zhu, Z.~Liu, C.~Miao, F.~Hajiaghajani, L.~Su, and C.~Qiao, ``Towards backdoor attacks against lidar object detection in autonomous driving,'' in \emph{Proceedings of the 20th ACM Conference on Embedded Networked Sensor Systems}, 2022, pp. 533--547.

\bibitem{chaturvedi2024badfusion}
S.~S. Chaturvedi, L.~Zhang, W.~Zhang, P.~He, and X.~Yuan, ``Badfusion: 2d-oriented backdoor attacks against 3d object detection,'' \emph{arXiv preprint arXiv:2405.03884}, 2024.

\bibitem{ma2022dangerous}
H.~Ma, Y.~Li, Y.~Gao, A.~Abuadbba, Z.~Zhang, A.~Fu, H.~Kim, S.~F. Al-Sarawi, N.~Surya, and D.~Abbott, ``Dangerous cloaking: Natural trigger based backdoor attacks on object detectors in the physical world,'' \emph{arXiv preprint arXiv:2201.08619}, 2022.

\bibitem{ma2024watch}
H.~Ma, S.~Wang, Y.~Gao, Z.~Zhang, H.~Qiu, M.~Xue, A.~Abuadbba, A.~Fu, S.~Nepal, and D.~Abbott, ``Watch out! simple horizontal class backdoor can trivially evade defense,'' in \emph{Proceedings of the 2024 on ACM SIGSAC Conference on Computer and Communications Security}, 2024, pp. 4465--4479.

\bibitem{lan2023influencer}
H.~Lan, J.~Gu, P.~Torr, and H.~Zhao, ``Influencer backdoor attack on semantic segmentation,'' \emph{arXiv preprint arXiv:2303.12054}, 2023.

\bibitem{mao2023object}
J.~Mao, Y.~Qian, J.~Huang, Z.~Lian, R.~Tao, B.~Wang, W.~Wang, and T.~Yao, ``Object-free backdoor attack and defense on semantic segmentation,'' \emph{Computers \& Security}, p. 103365, 2023.

\bibitem{chen2022effective}
W.~Chen, B.~Wu, and H.~Wang, ``Effective backdoor defense by exploiting sensitivity of poisoned samples,'' \emph{Advances in Neural Information Processing Systems}, vol.~35, pp. 9727--9737, 2022.

\bibitem{huang2022backdoor}
K.~Huang, Y.~Li, B.~Wu, Z.~Qin, and K.~Ren, ``Backdoor defense via decoupling the training process,'' \emph{arXiv preprint arXiv:2202.03423}, 2022.

\bibitem{weber2023rab}
M.~Weber, X.~Xu, B.~Karla{\v{s}}, C.~Zhang, and B.~Li, ``Rab: Provable robustness against backdoor attacks,'' in \emph{2023 IEEE Symposium on Security and Privacy (SP)}.\hskip 1em plus 0.5em minus 0.4em\relax IEEE, 2023, pp. 1311--1328.

\bibitem{liu2018fine}
K.~Liu, B.~Dolan-Gavitt, and S.~Garg, ``Fine-pruning: Defending against backdooring attacks on deep neural networks,'' in \emph{International Symposium on Research in Attacks, Intrusions, and Defenses}.\hskip 1em plus 0.5em minus 0.4em\relax Springer, 2018, pp. 273--294.

\bibitem{wu2021adversarial}
D.~Wu and Y.~Wang, ``Adversarial neuron pruning purifies backdoored deep models,'' \emph{Advances in Neural Information Processing Systems}, vol.~34, pp. 16\,913--16\,925, 2021.

\bibitem{zheng2022data}
R.~Zheng, R.~Tang, J.~Li, and L.~Liu, ``Data-free backdoor removal based on channel lipschitzness,'' in \emph{European Conference on Computer Vision}.\hskip 1em plus 0.5em minus 0.4em\relax Springer, 2022, pp. 175--191.

\bibitem{zheng2022pre}
------, ``Pre-activation distributions expose backdoor neurons,'' \emph{Advances in Neural Information Processing Systems}, vol.~35, pp. 18\,667--18\,680, 2022.

\bibitem{zhao2020bridging}
P.~Zhao, P.-Y. Chen, P.~Das, K.~N. Ramamurthy, and X.~Lin, ``Bridging mode connectivity in loss landscapes and adversarial robustness,'' in \emph{International Conference on Learning Representations}, 2020.

\bibitem{zheng2023judging}
L.~Zheng, W.-L. Chiang, Y.~Sheng, S.~Zhuang, Z.~Wu, Y.~Zhuang, Z.~Lin, Z.~Li, D.~Li, E.~Xing \emph{et~al.}, ``Judging llm-as-a-judge with mt-bench and chatbot arena,'' \emph{Advances in Neural Information Processing Systems}, vol.~36, pp. 46\,595--46\,623, 2023.

\bibitem{achiam2023gpt}
J.~Achiam, S.~Adler, S.~Agarwal, L.~Ahmad, I.~Akkaya, F.~L. Aleman, D.~Almeida, J.~Altenschmidt, S.~Altman, S.~Anadkat \emph{et~al.}, ``Gpt-4 technical report,'' \emph{arXiv preprint arXiv:2303.08774}, 2023.

\bibitem{team2023gemini}
G.~Team, R.~Anil, S.~Borgeaud, J.-B. Alayrac, J.~Yu, R.~Soricut, J.~Schalkwyk, A.~M. Dai, A.~Hauth, K.~Millican \emph{et~al.}, ``Gemini: a family of highly capable multimodal models,'' \emph{arXiv preprint arXiv:2312.11805}, 2023.

\bibitem{saha2020hidden}
A.~Saha, A.~Subramanya, and H.~Pirsiavash, ``Hidden trigger backdoor attacks,'' in \emph{Proceedings of the AAAI conference on artificial intelligence}, vol.~34, no.~07, 2020, pp. 11\,957--11\,965.

\bibitem{pasquini2020trembling}
C.~Pasquini and R.~B{\"o}hme, ``Trembling triggers: exploring the sensitivity of backdoors in dnn-based face recognition,'' \emph{EURASIP Journal on Information Security}, vol. 2020, no.~1, p.~12, 2020.

\bibitem{abad2023sok}
G.~Abad, J.~Xu, S.~Koffas, B.~Tajalli, S.~Picek, and M.~Conti, ``Sok: A systematic evaluation of backdoor trigger characteristics in image classification,'' \emph{arXiv preprint arXiv:2302.01740}, 2023.

\bibitem{rossolini2023real}
G.~Rossolini, F.~Nesti, G.~D’Amico, S.~Nair, A.~Biondi, and G.~Buttazzo, ``On the real-world adversarial robustness of real-time semantic segmentation models for autonomous driving,'' \emph{IEEE Transactions on Neural Networks and Learning Systems}, 2023.

\bibitem{shen2019advspade}
G.~Shen, C.~Mao, J.~Yang, and B.~Ray, ``Advspade: Realistic unrestricted attacks for semantic segmentation,'' \emph{arXiv preprint arXiv:1910.02354}, 2019.

\end{thebibliography}

\end{document}